\documentclass[prd,english,american,aps,amsfonts,amsmath,tightenlines,nofootinbib,preprintnumbers,sort&compress,12pt]{revtex4}

\date{August 2, 2011} 

\def\ifundefined#1{\expandafter\ifx\csname#1\endcsname\relax}

\ifundefined{pdfoutput}
   \usepackage[dvips]{graphicx}
\else
   \usepackage[pdftex]{graphicx}
\fi
\usepackage{epstopdf}

\usepackage{hyperref}
\usepackage{hypernat}   

\catcode`@=11
\let\savesort=\NAT@sort@cites
\newcommand\nosort[1]{\edef\NAT@cite@list{#1}}
\def\citenosort#1{\let\NAT@sort@cites=\nosort \cite{#1}%
   \let\NAT@sort@cites=\savesort}
\catcode`@=12 

\usepackage[T1]{fontenc}
\usepackage[latin1]{inputenc}
\usepackage{amsmath}
\usepackage{amssymb}

\makeatletter
\usepackage{epsf}
\usepackage{graphicx}

\def\be{\begin{equation}}
\def\ee{\end{equation}}
\def\beq{\begin{equation}}
\def\eeq{\end{equation}}
\def\bea{\begin{eqnarray}}
\def\eea{\end{eqnarray}}
\def\bml{\begin{subequations}}
\def\blea{\bml\begin{eqnarray}}
\def\elea{\end{eqnarray}\end{subequations}}

\def\expect#1{\left\langle {#1} \right\rangle}
\def\half{{1 \over 2}}

\def\lcurl{\left\{}  

\usepackage{babel}
\makeatother
\begin{document}

\preprint{MIT-CTP-4284\qquad  SU-ITP-11/36}

\title{Eternal Inflation, Global Time Cutoff Measures, and a
Probability Paradox}

\author{Alan H. Guth}
\affiliation{Center for Theoretical Physics, Laboratory for Nuclear 
Science, and Department of Physics, \\ Massachusetts Institute of
Technology, Cambridge, MA 02139, USA}

\email{guth@ctp.mit.edu}

\author{Vitaly Vanchurin}
\affiliation{Stanford Institute for Theoretical Physics and Department of Physics,  Stanford University, Stanford,  CA 94305, USA}

\email{vanchurin@stanford.edu}

\selectlanguage{american}
\begin{abstract}
The definition of probabilities in eternally inflating universes
requires a measure to regulate the infinite spacetime volume, and
much of the current literature uses a global time cutoff for this
purpose.  Such measures have been found to lead to paradoxical
behavior, and recently Bousso, Freivogel, Leichenauer, and
Rosenhaus have argued that, under reasonable assumptions, the
only consistent interpretation for such measures is that time
must end at the cutoff.  Here we argue that there is an
alternative, consistent formulation of such measures, in which
time extends to infinity.  Our formulation begins with a
mathematical model of the infinite multiverse, which can be
constructed without the use of a measure.  Probabilities, which
obey all the standard requirements for a probability measure, can
then be defined by mathematical limits.  They have a peculiar
feature, however, which we call time-delay bias: if the outcome
of an experiment is reported with a time delay that depends on
the outcome, then the observation of the reports will be biased
in favor of the shorter time delay.  We show how the paradoxes
can be resolved in this interpretation of the measure.

\end{abstract}
\maketitle

\section{Introduction}

Much attention in recent years has been focussed on eternal
inflation
\cite{Steinhardt:1983, Vilenkin:1983xq, Linde:1986fd,
Linde:1993xx, Vanchurin:1999iv, Garriga:2005av, Vanchurin:2006xp,
Bousso:2006ev, Linde:2008xf, Garriga:2008ks, DeSimone:2008bq,
DeSimone:2008if, Noorbala:2010zy}, for at least two good reasons. 
The first is that almost all inflationary models seem to lead to
eternal inflation. It is possible to design models of inflation
that are not eternal, but at least to many of us, such models
look contrived.  While the word ``generic'' is not really
well-defined in this context, it nonetheless seems that inflation
generically becomes eternal.

The second reason stems from the observed fact that the expansion
of the universe is accelerating
\cite{Riess:1998cb,Perlmutter:1998np}, a fact which is explained
most simply by assuming a nonzero cosmological constant, or
vacuum energy density.  But the required value of the
cosmological constant is then more than 120 orders of magnitude
less than the Planck scale, a fact which is very hard to explain. 
For many of us, the most plausible explanation currently known is
the proposal, advanced by Weinberg in 1987, that the value of the
cosmological constant is governed by a selection effect
\cite{Weinberg:1987dv}.  If the idea of a string theory landscape
of vacua \cite{Bousso:2000xa, Kachru:2003aw, Kachru:2003sx,
Douglas:2003um, Susskind:2003kw, Linde:2009ah} is combined with
the assumption that eternal inflation produces an infinite number
of pocket universes which populate all or at least many of these
vacua, then the selection-effect argument is given a logical
setting: the many vacua of the landscape all have different
vacuum energy densities, with a large number (although a tiny
fraction) expected to be consistent with the observed value.  If
life forms preferentially in pocket universes with very small
values of the vacuum energy density, then the smallness of the
observed cosmological constant is explained.  Almost all life in
the multiverse would observe a very small value of the
cosmological constant.

Once eternal inflation is taken as a serious possibility, an
important question arises: how does one define probabilities in
an eternally inflating multiverse? In an eternally inflating
multiverse, anything that can happen will happen, an infinite
number of times.  Furthermore, in a world described by quantum
theory, almost anything can happen.  We normally think that
physical theories can give meaningful predictions because, on the
macroscopic level, some classes of outcomes are much more
probable than others.  But in a multiverse theory all these
outcomes will happen an infinite number of times, so a physical
theory can make useful predictions only if one can distinguish
between common events and very rare ones.  But to say that one
type of event is more probable than another, one has to compare
infinities.  Comparing infinities is not in general a
well-defined operation, so one needs to introduce some
prescription for regularizing the infinities.  Such a
prescription is called a measure.  With a measure to regularize
the infinities, it would be possible to say that one sequence of
events happens twice as often as some other sequence, and is
therefore twice as probable.  Without a measure, a theory could
give no information other than to distinguish possible events
from events which are totally impossible.

Global time cutoff measures, especially the scale-factor cutoff
measure \cite{DeSimone:2008if}, appear to be promising and have
recently received much attention.  The detailed definition of a
global time cutoff measure will be given in the next section. 
Global time cutoff measures can be constructed for a variety of
different time variables, and the consequences of the measure can
depend in a very crucial way on the choice of time variable. 
Nonetheless, the discussion here will focus only on qualitative
issues which are common to all global time cutoff measures, so we
will not need to specify what time variable is being used. 

In a recent paper by Bousso, Freivogel, Leichenauer, and
Rosenhaus \cite{Bousso:2010yn}, which we will refer to as BFLR, a
point of view is described in which the final time cutoff is
interpreted as an absolute cutoff, the literal end of time.  The
possibility of hitting the cutoff is described as a ``novel type
of catastrophe.'' As they put it, ``we argue that cutoff
observers are a real possibility, because {\it there is no
well-defined probability distribution without the cutoff; in
particular, only the cutoff defines the set of allowed events.}''
We will refer to this description as the absolute cutoff
interpretation.

In this paper we will present an alternative point of view, which
we will refer to as the mathematical limit interpretation.  In
this formulation the multiverse goes on forever with no end of
time, and the cutoff serves only as a mathematical device for
defining probabilities.  The cutoff literally disappears once the
limit $\tau_c \to \infty$ is taken.  We will of course not argue
that this type of measure is necessarily correct.  We do not
claim to know the correct answer to the measure question, and so
far as we know, nobody else does either.  Furthermore, one is
always free to assume that time will end if one wants to, so we
will not try to argue that this is not possible.  However, we
will argue that global time cutoff measures are logically
consistent, and that they are perfectly compatible with time
continuing without limit.  Eternal inflation, with a global time
cutoff measure, does not require an end of time.

In Sec.~II we will describe the definition of a global time
cutoff measure, starting with a fairly detailed description of
how one can define an infinite lattice model of the multiverse. 
An important feature of the mathematical limit interpretation is
that even though the multiverse is infinite, it can still be
given a mathematical definition that makes it in principle as
well-defined as the set of positive integers.  The measure is
needed to count events that occur in the multiverse, but the
multiverse itself can be constructed without a measure.  In
Sec.~III we discuss some peculiar properties of global time
cutoff measures: the youngness bias, the fact that a nonzero
fraction of all observers reach the cutoff (even as it is taken
to infinity), and a probability paradox that focuses on an
experiment in which the results of a coin flip are announced with
differing time delays, depending on the outcome.  The absolute
cutoff interpretation gives a straightforward resolution of this
paradox, but it is much more subtle in the mathematical limit
interpretation.  Before discussing the resolution, in Sec.~IV we
review some mathematical properties of probabilities defined on
infinite sets.  We return to the probability paradox in Sec.~V,
showing that it can be resolved by recognizing a
counter-intuitive feature of global time cutoff measures:
time-delay bias.  That is, if the outcome of an experiment is
reported with a time delay that depends on the outcome, then the
observation of the reports will be biased in favor of the shorter
time delay.  We go on to discuss betting on the paradoxical
experiment, which we treat as a consistency test of our
understanding of the probabilities. We discuss in particular the
possibility of bets for which the time of the payoff depends upon
the outcome.  The time-delay bias suggests that maybe one can
actually influence the results of an experiment by introducing an
outcome-dependent time delay, so we discuss this issue in
Sec.~VI.  We summarize our conclusions in Sec.~VII.  In an
appendix, we attempt to show that the time-delay bias can be
understood in terms of the cloning --- i.e., the appearance of an
exponentially growing set of copies of any experiment --- that is
a distinctive feature of eternal inflation.

\section{Definition of Global Time Cutoff Measures}
\label{sec:definition}

The first step in this mathematical limit interpretation is to
recognize that the entire, infinite multiverse can be constructed
(mathematically) {\it before} the probability measure is even
mentioned.  But how can one imagine ``constructing'' an infinite
system? The method is essentially the same as the one that has
been used by mathematicians since the 1800's.  In 1889 Giuseppi
Peano published the famous set of axioms that are now
traditionally taken as the basis for the positive integers
\cite{Peano}.  The infinite system is constructed by postulating
that the number one (or perhaps zero) exists, and that each
natural number has a successor.  A few other axioms are
introduced to assure that each application of the successor
function produces a number that is unequal to all previous
numbers, and the infinite set is constructed.

It is worth noting that infinite sets, such as the set of
positive integers, are {\it not} constructed as the limit of
finite sets.  While at first it might look sensible to define the
set of all integers as the set of integers from 1 to $N$, in the
limit as $N \to \infty$, a little thought shows that this is not
consistent with the usual concept of a limit. Consider, for
example, the concept of a limit in the context of a real-valued
function $f(x)$ of a real number $x$.  The limit
\beq
  \lim_{x \to x_0} f(x) = a 
\label{eq:limit}
\eeq
is defined to be true if and only if for every real $\epsilon>0$
there exists a real $\delta > 0$ such that $0 < |x-x_0| < \delta$
implies that $|f(x) - a| < \epsilon$.  The crucial point is that
$f(x)$ can be said to approach $a$ only if the distance between
$f(x)$ and $a$ can become arbitrarily small as $x$ becomes closer
and closer to $x_0$.  But if we try to apply the same notion to
sets, we see that the logic does not carry over, since there is
no sense in which the distance between a finite set and an
infinite set is ever small.  The set of integers from 1 to $N$ is
always infinitely different from the set of all integers, so it
does not approach the set of all integers as $N$ approaches
infinity.  Thus, infinite sets are not constructed as limits, but
rather by giving a rule, such as the Peano successor axiom, that
determines the elements of the set.

So, just as the infinite set of positive integers is constructed
by assuming the successor rule of the Peano axioms, a model of
the infinite multiverse can be constructed by its update rule,
which is determined by the model of evolution embodied in the
laws of physics.  For purposes of discussing the measure we
pretend that the laws of physics are known, so that we can focus
on the problem of how probabilities are defined.  To think
concretely, we will also assume that these laws of physics can be
well-approximated on a discrete lattice, where we will refer to
each lattice site as a pixel.

The choice of time variable is a crucially important element of
any lattice simulation, but here we are not aiming for
efficiency.  We are merely trying to give a simple description of
how such a simulation can in principle be carried out.  For this
purpose we can imagine using proper time $t$.  The 3-volume
$V_{\rm multiverse}(t)$ of the multiverse is expected to grow
exponentially at late times, with some rate that we will call
$\lambda$.  That is,
\beq
V_{\rm multiverse}(t) \propto e^{\lambda t} \ .
  \label{eq:volume}
\eeq
(One might imagine a multiverse that is infinite in volume at all
times, but it suffices for our purposes to consider a multiverse
model that is finite at any given time, but which continues to
grow for an infinite time.  This provides an infinite spacetime
volume that can be sampled to define probabilities.) We believe
that this formula for the volume is exact in the limit of large
$t$, depending only on the assumption that the late-time
evolution becomes self-similar.  That is, at asymptotically late
times we can think of the multiverse as a huge number of cosmic
regions, each including a huge number of pocket universes.  Each
of these regions is statistically identical to the others and
evolves independently of the others.  Since we can choose these
regions as large as we like, we can presumably arrange for
statistical fluctuations to be as small as we like. Thus, in the
amount of time $t_{\rm e\hbox{\scriptsize -}fold}$ in which one
region enlarges by a factor of $e$, so will all the other
regions.  And in the next increment $t_{\rm e\hbox{\scriptsize
-}fold}$
of time, all regions will again enlarge by a factor of $e$.%
\footnote{We assume (but do not prove) the
self-similarity of the evolution, and argue that self-similarity
implies that asymptotically the growth is precisely exponential. 
The argument can perhaps be phrased more clearly in terms of the
lattice model.  At some very late time, suppose that the lattice
consists of $Q^6$ sites, for some very large $Q$ (say $Q \sim 1$
googol).  Suppose further that it can be divided into $Q^3$ cubic
regions, each of size $Q \times Q \times Q$ pixels, and each
containing a huge number of pocket universes.  By the homogeneity
of the construction, any two of these regions would be
statistically identical, so the expectation value of any quantity
should be the same for the two regions.  Furthermore, the
statistical fluctuations of any quantity, when expressed as a
fraction of the mean, will appoach zero as $Q \to \infty$.  We
can imagine calculating the expansion factor of each region
during the next time step $\Delta t$, and the expectation values
for the two regions should be equal, with the fluctuations
arbitrarily small as $Q \to \infty$.  Now consider the state of
the lattice at a much later time, focusing on any cubic region of
the lattice with the same physical size, in the sense that the
expectation value for its 3-volume matches that of the original
regions.  Our assumption of self-similarity includes the
assumption that any such region should also be statistically
identical to the original regions, and therefore should have the
same expectation value for the expansion factor during the next
time interval $\Delta t$, again with arbitrarily small
fluctuations.  Note that all these expecation values are
well-defined before we introduce a measure, since they are
defined by the probabilities for fields at specific lattice
sites.  Under these assumptions, the expansion factor for the
time interval $\Delta t$ can be converted into an exponential
expansion rate $\lambda$, which describes the expansion of the
entire multiverse.}

\begin{figure}[htbp]
   \begin{center}
   \includegraphics[width=3.5in]{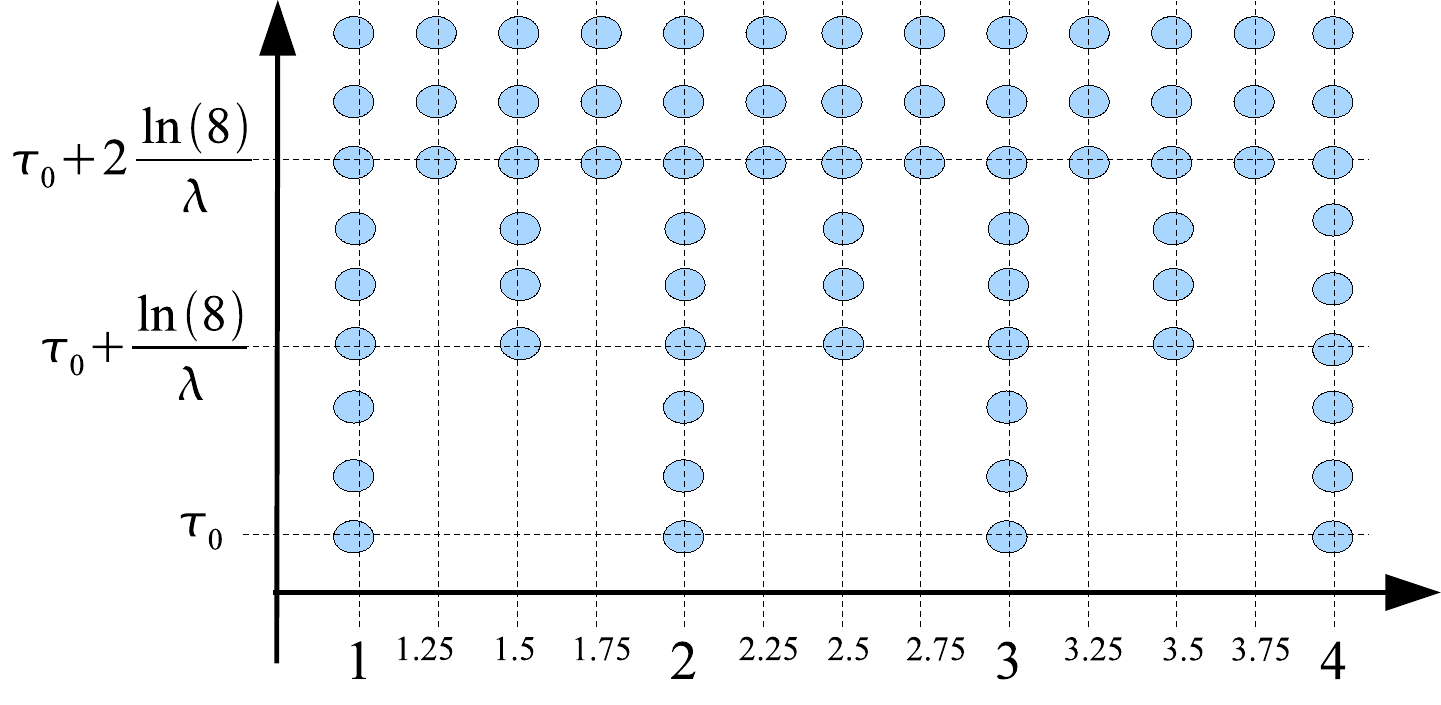}
 \caption{Pixelated multiverse.}
   \label{fig:pixelated}
   \end{center}
   \end{figure}

To describe the exponentially expanding multiverse on the
lattice, we can imagine that the number of pixels at a given
value of the time coordinate increases exponentially with the
time, so the average lattice spacing does not change with time.
More precisely, to describe an eternally inflating multiverse
with late time exponential growth of 3-volume given by
Eq.~(\ref{eq:volume}), we can introduce a doubling of the lattice
points in each direction whenever the total 3-volume increases by
a factor of $2^3=8$. Then the series of mesh refinement times
would be given by an arithmetic series $\{\tau_0, \tau_0+{\ln(8)
\over\lambda}, \tau_0+2 {\ln(8)\over\lambda}, \tau_0+3
{\ln(8)\over\lambda},...\} $ with the total number of lattice
points increasing by a factor of 8 at each coordinate time listed
in the series. (See Fig.\ \ref{fig:pixelated}%
).

We will not need to fully describe the details of this model, but
will assume that the multiverse can be described by specifying
the values of some number of fields, including the metric and
various particle fields, at each lattice site.  For simplicity,
we assume that even the set of possible field values can be
discretized on some fine mesh.  We want to focus on the treatment
of the infinite spacetime volume and the infinite number of
events that will result, so we try to simplify the local
description of the physics as much as possible.  As long as the
discretizations of the spacetime lattice and the values of the
fields are chosen on a very fine mesh, we assume that the local
laws of physics can be approximated to as high an accuracy as we
wish.  In any case, we assume that the discretization of the
local laws of physics does not affect the fundamental issues that
arise from the infinite spacetime volume.

A model of the multiverse can then be constructed by starting
with some finite-sized initial state, defined on a finite number
of pixels, and letting the system evolve.  At each step of the
(discretized) evolution, the update rule will define the joint
probability for every possible set of pixel values in terms of
the probabilities that were known for the previous time step.  We
assume that general coordinate invariance and any gauge
symmetries of the fields can be handled by some gauge-fixing
procedure, so the evolution becomes uniquely defined.  We further
assume that even processes like bubble nucleation can be
described on the lattice if the lattice spacing is small enough,
and that higher-dimensional physics with compactification could
be simulated on a lattice if necessary.  To allow for final
singularities in black holes and in
negative-cosmological-constant bubbles, the field values of a
pixel should include the option of the pixel being nonexistent.
In any case, once the update rule is specified, the multiverse
model is fully constructed, and is just as mathematically
well-defined as the set of positive integers.  Note that the
probabilities for the pixelated multiverse are calculated
directly from the assumptions about the initial conditions and
the rules of evolution.  The measure is needed to count events
within the multiverse, but the description of the multiverse
itself does not require a measure.

Note that if we attempted to implement these update rules on a
computer, we would only be able to proceed for some finite number
of steps before the simulation exceeded the size of the computer. 
But we view this as a practical problem, not a problem in
principle.  If we programmed a computer to list the integers,
there would also be a limit, at which point the computer would
run out of registers.  But the limitations of a computer are
never interpreted as implying an end to the integers.  The set of
integers is mathematically well-defined, and infinite. 
Similarly, the model of the multiverse would be well-defined, and
infinite, even if any specific computer is limited to describing
a finite piece of it.

In this mathematical limit interpretation, we see that it is not
true that the cutoff defines the set of allowed events, as was
assumed in the absolute cutoff interpretation.  Rather the full,
infinite set of events is constructed as a mathematical object
before the cutoff is even mentioned.  It is the laws of physics,
or our best approximation of them, that defines the set of
allowed events.

\begin{figure}[htbp]
   \begin{center}
   \includegraphics[width=3.5in]{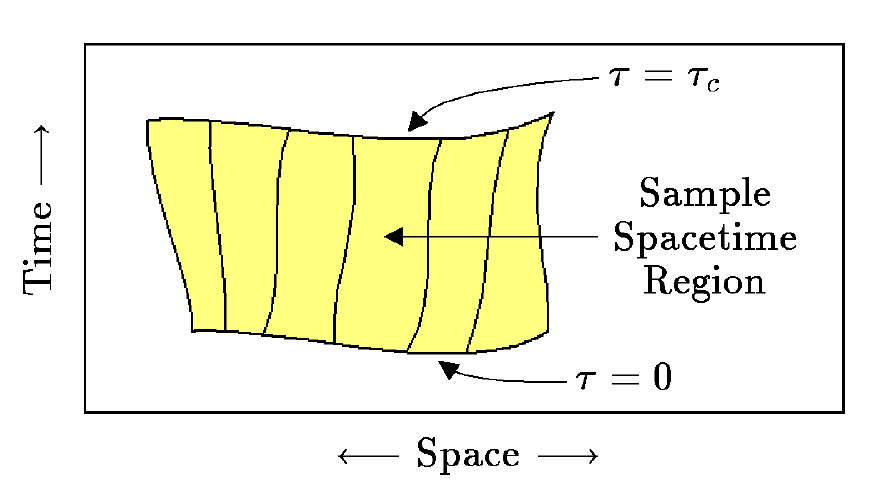}
   \caption{Global time cut-off measure.}
   \label{fig:measure2}
   \end{center}
   \end{figure}

Once the mathematical model of the multiverse has been
constructed, a measure can be introduced to define the relative
abundance of different kinds of events.  We are working in the
semiclassical approximation, so we describe the spacetime
classically, with a stochastic evolution that reflects the
underlying quantum dynamics.  The construction starts by making
an arbitrary choice of an initial, finite, spacelike slice, which
is described as $\tau=0$, where $\tau$ is a time variable used in
the definition of the measure.  It could be the same proper time
variable $t$ that was used in the lattice simulation, but it
could also be scale-factor time or some other time variable. (See
Fig.\ \ref{fig:measure2}%
).  One then constructs a family of timelike geodesics, each
normal to the initial surface, projecting them into the future. 
The time variable $\tau$ is measured along each geodesic, with a
definition determined by the choice of time variable.  The
worldlines are each followed until they reach an arbitrary final
time cutoff $\tau=\tau_c$. The region swept out by these
worldlines, between $\tau=0$ and $\tau=\tau_c$, becomes the
sample spacetime region.  There is a minor complication if the
construction does not encounter a region of eternal inflation,
which can happen for some choices of the initial (finite)
spacelike slice.  If that happens, the recipe requires that a new
spacelike slice is chosen, and the construction starts again. 
Once a region of eternal inflation is found, however, the sample
spacetime region will approach infinite spacetime volume as
$\tau_c \to \infty$. The relative abundance of any two types of
events $A$ and $B$ is then defined as the ratio of the numbers of
these events in the sample spacetime region, $N_A/N_B$, in the
limit as $\tau_c$ approaches infinity.  The limit is expected to
be independent of the choice of initial surface, and in this
measure it defines the relative abundance and the relative
probability of these events:
\beq
{P_A \over P_B} \equiv \lim_{\tau_c \to \infty} \, {N_A \over
     N_B} \ .
\label{eq:events}
\eeq

In the paragraph above we talked about ``types of events,''
without clarifying what exactly we meant.  The phrase ``type of
event'' should mean a class of events that is described with
well-defined tolerances, so that if anybody looked at what was
happening in a region of spacetime, she could decide without
ambiguity whether an event of this class has taken place. 
Examples of ``types of events'' would include the birth of normal
observers, the nucleation of Boltzmann brains
\citenosort{Dyson:2002pf, Albrecht:2004ke, Page:2005ur,
DeSimone:2008if} that are smart enough to simulate normal
observers for at least a second (assuming that ``normal
observer'' has been suitably defined), the decays of protons,
etc.

In addition to events, which are occurrences that can be
approximated as points in spacetime, we may also want to consider
long-term projects, like a 10-year-long proton decay experiment. 
For the proton decay experiment, one can talk about the pushing
of the upload button that sends the final paper to the arXiv ---
which is well-approximated as a point-like event --- but one
might want to also talk about probabilities for the full
experiment.  If we look only at the pushing of upload buttons, we
would be counting bad experiments and possibly even fraudulent
experiments, while if we include the full experiment in our
description, we could in principle devise a definition that
excludes such things.  We will call such extended items
``stories''.  So, a type of story would be defined as a
description of a finite-sized region of spacetime that is
specified with well-defined tolerances, so that if anybody looked
at what was happening in a region of spacetime, she could decide
without ambiguity whether or not this story occurs in the region. 
In the global time cutoff measure, the relative probabilities of
any two types of stories $A$ and $B$ are again given by
Eq.~(\ref{eq:events}), where now $A$ and $B$ are generalized to
refer to types of stories. We will consider an event to be a
special (point-like) case of a story, so Eq.~(\ref{eq:events})
covers all cases.

\begin{figure}[htbp]
   \begin{center}
   \includegraphics[width=3.4in]{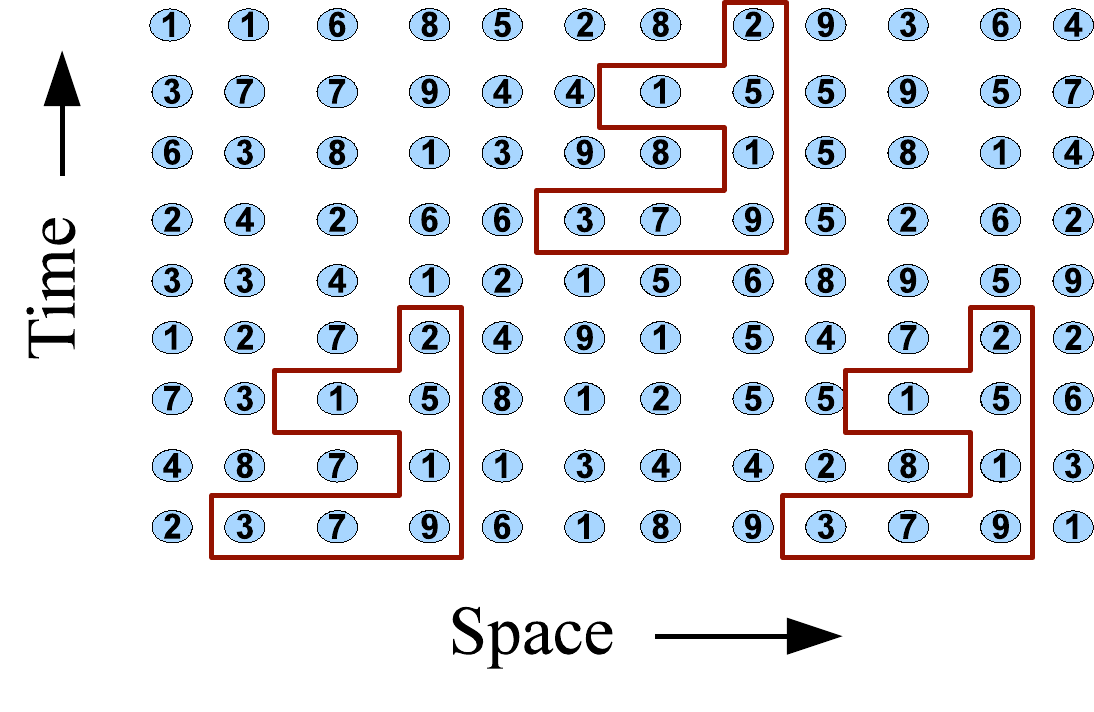}
   \caption{Microstories in a 
(1+1)-dimensional pixelated multiverse.}
   \label{fig:microstory}
   \end{center}
\end{figure}

To give a precise definition of a story in the context of the
lattice model, we begin by defining a microstory as a story with
a complete description of a region where the microstory is
constructed. In general the region does not have to be simply
connected nor connected, but it does have to be finite.  An
example of a microstory is shown in Fig.\ \ref{fig:microstory}
in the context of a (1+1)-dimensional pixelated multiverse. The
figure contains only three occurrences of the same microstory
encircled by red lines, but, of course, in an infinite multiverse
each microstory, which is not forbidden by laws of physics, would
occur an infinite number of times.  To describe a given
microstory precisely we can start by first defining its origin,
and then we can list the relative coordinates of all of its
pixels with their exact values. The origin of a story can be
defined by identifying all of its pixels with the smallest
$\tau$, out of which one can identify all of the pixels with the
smallest $x$, and so on for $y$ and $z$, which will always give a
unique pixel. Then for the particular example of Fig.\
\ref{fig:microstory}, the microstory $M$ is defined by the set of
relative coordinates and pixel values:
\beq
M \equiv \{(0, 0; 3), (0,1;7), (0, 2; 9), (1, 2; 1),
     (2,1;1),(2,2;5), (3,2;2)\} .
\eeq
If the microstory overlaps one or more mesh refinements, then the
microstory description must specify the times at which they
happen; after a mesh refinement the spatial coordinates are
allowed to have half-integral values, with further subdivisions
if there are more mesh refinements.

For a given realization of the multiverse lattice, a microstory
is said to match at a given lattice site if all the pixels in $M$
match when the origin of $M$ is assigned to the lattice site. 
(In matching, the pixel specification $(0,1;7)$ is always
compared with the pixel one to the right of the origin, even if
the mesh-refined coordinate increment on the lattice is 1/2, 1/4,
or anything else.) The match is in the sample spacetime region if
all of its pixels are.  We can let $N(M)$ denote the number of
matches for the microstory $M$ in the sample spacetime region.

A story $S$ is then defined as a finite set of distinct
microstories
\beq
S = \{M_1, M_2, \dots, M_n\} \ . 
  \label{eq:storydef}
\eeq
Any such set can be called a story, but we will mainly be
interested in stories that correspond to a macroscopic
description, such as a coin toss landing heads.  Ideally the set
$\{M_1, M_2, \dots, M_n\}$ can be constructed so that every
situation that would macroscopically be described as a coin toss
landing heads would match one and only one microstory in the set.
It would not be easy to construct such a set, but we assume that
it can be done.  In many cases the macroscopic description might
not specify an orientation, so the set of microstories would have
to include approximations to spatially rotated states, even
though the lattice breaks rotational symmetry microscopically. 
Time translations are also complicated, because the set of
microstories should include all possible ways that the
macroscopic version of the story can be cut by mesh refinements,
and all possible combinations of pixel sizes and metric values.
As long as we are convinced that this can be done, however, we do
not need to think about the details.

We then define the number of matches for the story $S$, in the
sample spacetime region, by summing the microstories:
\beq
N(S) = \sum_{j=1}^n N(M_j) \ .
  \label{eq:storysum}
\eeq

Given two stories $S_A$ and $S_B$, it is useful to define their
union, $S_A \cup S_B$, which corresponds intuitively to a story
which fits the description of either $S_A$ or $S_B$.
Mathematically, the union can be constructed by taking the union
of the sets of microstories:
\beq
S_A \cup S_B = \{M_1^A, M_2^A, \dots, M_{n_A}^A\} \cup \{M_1^B,
     M_2^B, \dots, M_{n_B}^B\} \ .
\label{eq:unions}
\eeq
Two stories $S_A$ and $S_B$ can be said to be storywise disjoint
if there is no possible pixel assignment that can match both of
the stories at the same lattice site.  (Note that this is a
stronger condition than requiring $\{M_1^A, M_2^A, \dots,
M_{n_A}^A\}$ to be disjoint from $\{M_1^B, M_2^B, \dots,
M_{n_B}^B\}$.) If $S_A$ and $S_B$ are storywise disjoint, it is
easy to see that
\beq
N(S_A \cup S_B) = N(S_A) + N(S_B) \ .
\label{eq:additivity}
     \eeq

We are mainly interested in defining probabilities for the
outcomes of experiments, so we need to define in detail what an
experiment is.  Intuitively, an experiment is when some system is
either constructed or perhaps found, and then the system is
watched for some finite amount of time to see how it develops. 
Generally there are a discrete number of possible outcomes, where
the discreteness may come about by binning the data.

An experiment can be described precisely in terms of stories by
constructing, for each possible outcome $i$, a story $S_i$ which
includes the setup of the experiment and the outcome.  We assume
that the setup is common to all the outcomes.  We also assume
that $S_i$ is in all cases storywise disjoint from $S_j$, since
the outcomes are distinguishable.  We can define a story for the
experiment, $S_{\rm expt}$ as the union of the $S_i$,
\beq
S_{\rm expt} = \bigcup_i S_i \ . 
  \label{eq:union}
\eeq
Then, based on the generic probability formula of
Eq.~(\ref{eq:events}), we can write the probability for the
outcome $i$ as
\beq
P(i) = \lim_{\tau_c \to \infty} {N(S_i) \over N(S_{\rm expt})} \
.
  \label{eq:expt}
\eeq
Given Eq.~(\ref{eq:additivity}), it is clear that these
probabilities sum to 1, giving a well-defined probability space
consistent with all the standard requirements.

\section{Oddities of Global Time Cutoff Measures}

There are several features that arise in global time cutoff
measures that appear at first to be surprising, and in particular
we would like to discuss three.

\subsection{Youngness bias}
\label{sec:youngness}

For the same reason that we expect the 3-volume of the multiverse
to expand exponentially with time at late times, as in
Eq.~(\ref{eq:volume}), the 4-volume $V_{\rm SSR}$ of the sample
spacetime region can also be expected to grow exponentially with
the cutoff time variable $\tau$.  Since it may be a different
time variable from proper time, the exponential expansion rate
with respect to $\tau$ might be different, so we call it
$\lambda_c$:\footnote{For scale factor time there is a further
complication, as discussed in Ref.~(\cite{Garriga:2005av}).  In
this case the volume expansion factor is always $e^{3t}$, by the
definition of $t$, so the regions that dominate the spacetime
volume at late times are those regions with the slowest decay
rate.  Thus the spacetime volume at late times is dominated by
stable (supersymmetric) Minkowski vacua, which grow as $e^{3t}$. 
The stories which we want to count, however, occur in inflating
vacua (with positive energy densities), or in the early stages of
terminal vacua (with zero or negative energy densities).  The
spacetime volume of this region grows with a subleading
exponential behavior, $e^{\lambda_c \tau_c}$, where $\lambda_c$
is slightly less than 3.  It is this spacetime volume that is
relevant to counting stories in the multiverse.}
\beq
V_{SSR} \propto e^{\lambda_c \tau_c} \ .
  \label{eq:SSR}
\eeq
The youngness bias is a consequence of this exponential growth,
which implies that most of the spacetime volume in the SSR lies
within a few time constants $1/\lambda_c$ from the final cutoff
$\tau=\tau_c$.  If we imagine that $1/\lambda_c$ is a short time,
then most of the pocket universes in the SSR are very young, with
ages less than a few times $1/\lambda_c$. Thus, the distribution
of pocket universes in the SSR is strongly dominated by very
young ones, with older pocket universes being very rare.  For
proper-time cutoff measure, $1/\lambda_c$ might be of order
$10^{-38}$ second (if the fastest inflation happens at the grand
unified theory scale), so the youngness bias is absurdly strong
\cite{Guth:2000ka}.  The proper-time cutoff measure can be ruled
out on this basis.  (For example, Max Tegmark pointed out that if
proper-time cutoff measure governed probabilities in the real
universe, then the probability that we would observe a cosmic
microwave background temperature lower than 3 K would be
$10^{-10^{56}}$~\cite{Tegmark:2004qd}. It is overwhelmingly more
likely for the universe to be younger, and hence hotter.) For
scale-factor cutoff measure $1/\lambda_c$ is of the order of the
Hubble time, so the youngness bias is extremely mild. 
Nonetheless the youngness bias applies in principle to the
scale-factor cutoff measure, so the conceptual issues that the
youngness bias raises are relevant for questions of
interpretation and logical consistency, although quantitatively
the effects would be undetectable.

\subsection{Nonzero fraction of all observers reach the cutoff}

The exponential expansion also leads to a peculiar result
concerning the distribution of observers.  For any given value of
the cutoff time $\tau_c$, a nonzero fraction of all observers who
have ever lived are still alive at the cutoff.  This fraction
approaches a nonzero constant in the limit that $\tau_c \to
\infty$.  The interpretation of this fact will be one of the main
concerns of this paper. 


In the absolute cutoff approach of BFLR, the cutoff is taken as a
genuine end of time at some large but finite $\tau_{\rm end}$,
and observers who reach the cutoff cease to exist at that point.
The calculated probabilities then depend only on the multiverse
at times $\tau < \tau_{\rm end}$, and times later than $\tau_{\rm
end}$ are meaningless.  In the mathematical limit interpretation,
it is really the opposite.  Probabilities of events, or more
generally stories, are defined by the limit in
Eq.~(\ref{eq:events}), which is completely controlled by
arbitrarily late times $\tau$.  For any value $\tau_{\rm chosen}$
that one might choose to consider, one could imagine making
arbitrary changes in the probabilities for $\tau <
\tau_{\rm chosen}$, and then recalculating the limit.  Since the
limit is determined by the behavior of the probabilities as $\tau
\to \infty$, the changes made for $\tau < \tau_{\rm chosen}$ have
no effect whatever.  So, for the absolute cutoff interpretation,
the probabilities are completely determined by the region $\tau <
\tau_{\rm end}$.  In the mathematical limit interpretation, the
probabilities are determined solely by the behavior of the
multiverse for $\tau > \tau_{\rm chosen}$, for any (finite)
choice of $\tau_{\rm chosen}$. 

\begin{figure}[htbp]
   \begin{center}
   \includegraphics{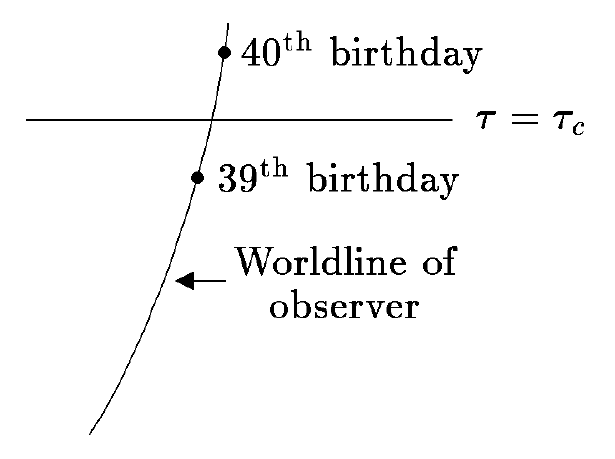}
   \caption{Cut-off surface intersected by the world-line of a
thirty-nine-year old.}
   \label{fig:birthdays}
   \end{center}
   \end{figure}

In implementing Eq.~(\ref{eq:events}), however, we have to be
able to count stories for a fixed value of $\tau_c$, before we
take the limit.  How, then, should we interpret the worldline of
an observer who, for example, reaches the cutoff between her 39th
and 40th birthdays, as shown in Fig.\ \ref{fig:birthdays}%
? In the absolute cutoff interpretation, this is a person who
suffers the catastrophe of reaching the end of time at age 39. 
In the mathematical limit interpretation, however, there is no
such catastrophe. The multiverse continues beyond the cutoff, and
even the calculation of probabilities will continue beyond the
cutoff at the next stage, as the cutoff is taken to infinity. 
So, following Eq.~(\ref{eq:events}) and the definition of a
story, such a worldline is counted in the class of stories of
observers who have had their 39th birthdays, but not in the class
of stories of observers who have had their 40th birthdays.  A
person meeting this description is simply a 39-year-old person. 
In the exponentially expanding multiverse there is a finite ratio
between the number of currently living 39-year-olds and the total
number of observers who have ever lived.  A worldline like the
one shown in Fig.\ \ref{fig:birthdays} is a contribution to the
numerator of this ratio.

\subsection{The ``Guth--Vanchurin'' Paradox}
\label{sec:Guth-Vanchurin}

In their justification for the absolute cutoff interpretation,
BFLR explain that it provides a clear resolution for a paradox
that we had brought up in private discussions.  Following BFLR,
we will refer to this as the G-V paradox.  Here we describe a
slight variant of that paradox, which contains the same relevant
features but which is slightly easier to analyze.

\begin{figure}[htbp]
   \begin{center}
   \includegraphics[width=3.5in]{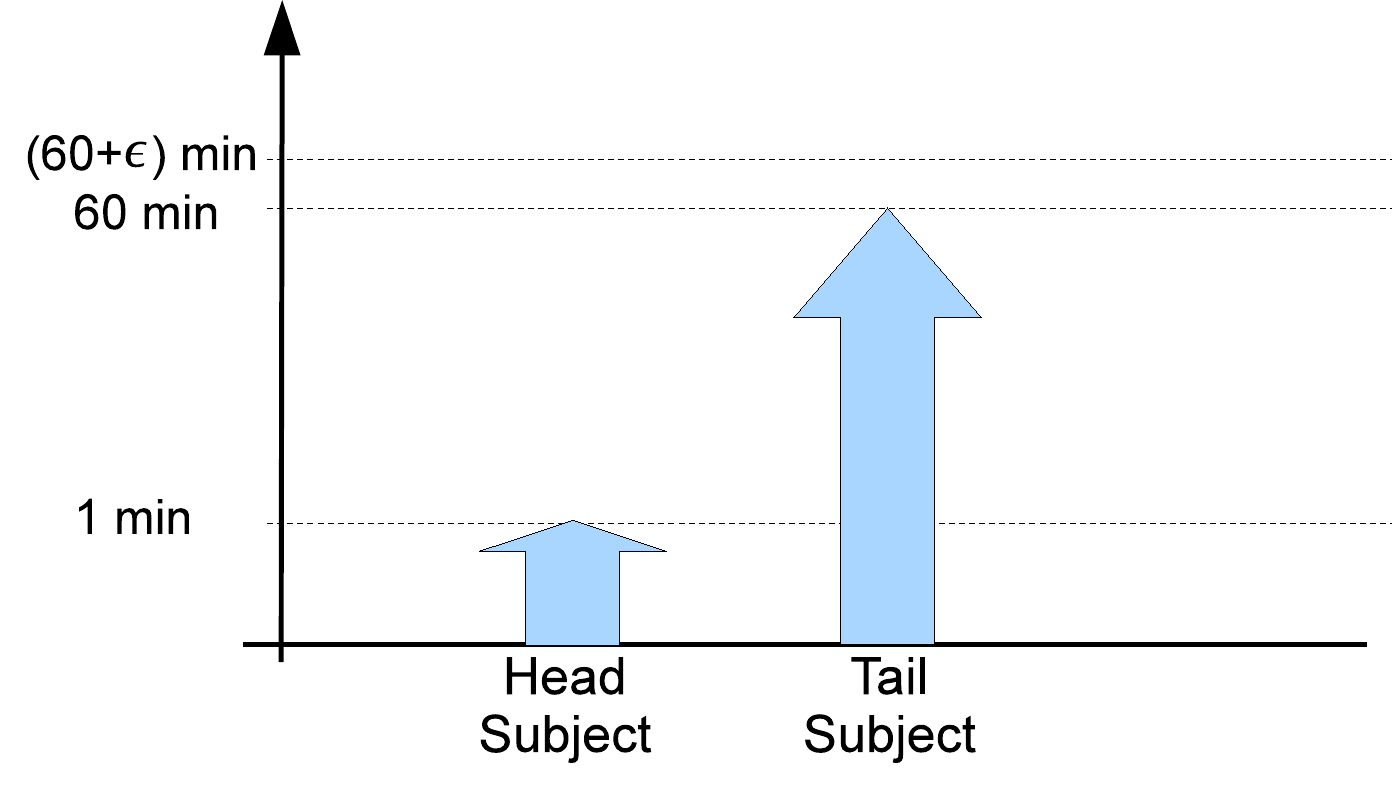}
   \caption{Sleeping periods in the G-V experiment.}
   \label{fig:subjects}
   \end{center}
   \end{figure}

This version of the paradox involves a thought experiment with
one experimenter and two subjects, Subject 1 and Subject 2.  The
experimenter flips a fair coin, but does not show the result to
the subjects.  The subjects then both go to sleep.  If the coin
was a head, then Subject 1 is designated as the Head Subject, and
Subject 2 is designated as the Tail Subject.  If the coin was a
tail, the designations are reversed.  The experimenter then sets
an alarm clock for each subject.  The Head Subject's clock is set
for a short nap, of length $\Delta t_{\rm short}$, and the Tail
Subject's clock is set for a long nap, of length $\Delta t_{\rm
long}$.  For simplicity of language, we will refer to $\Delta
t_{\rm short}$ as one ``minute,'' and $\Delta t_{\rm long}$ as
one ``hour'' (See Fig.\ \ref{fig:subjects}%
).  We will refer to $\Delta t_{\rm long} - \Delta t_{\rm short}$
as ``59 minutes.'' We assume that the universe expands by a
factor $Z$ during the 59 minutes between the two intervals, where
$Z$ is distinguishable from one.  In fact, for purposes of
discussion, we will adopt the assumption that $Z \gg 1$.  We will
also assume for simplicity that $Z$ has the same value everywhere
in the multiverse.\footnote{This version of the paradox differs
from the original by introducing the second subject.  In this
version the short and long nappers are not merely statistically
equal in number, but are paired one to one, which allows a
slightly simpler discussion. From the point of view of either
subject, however, it is no different from the original version.}

\begin{figure}[htbp]   
      \begin{center}
   \includegraphics[width=3.5in]{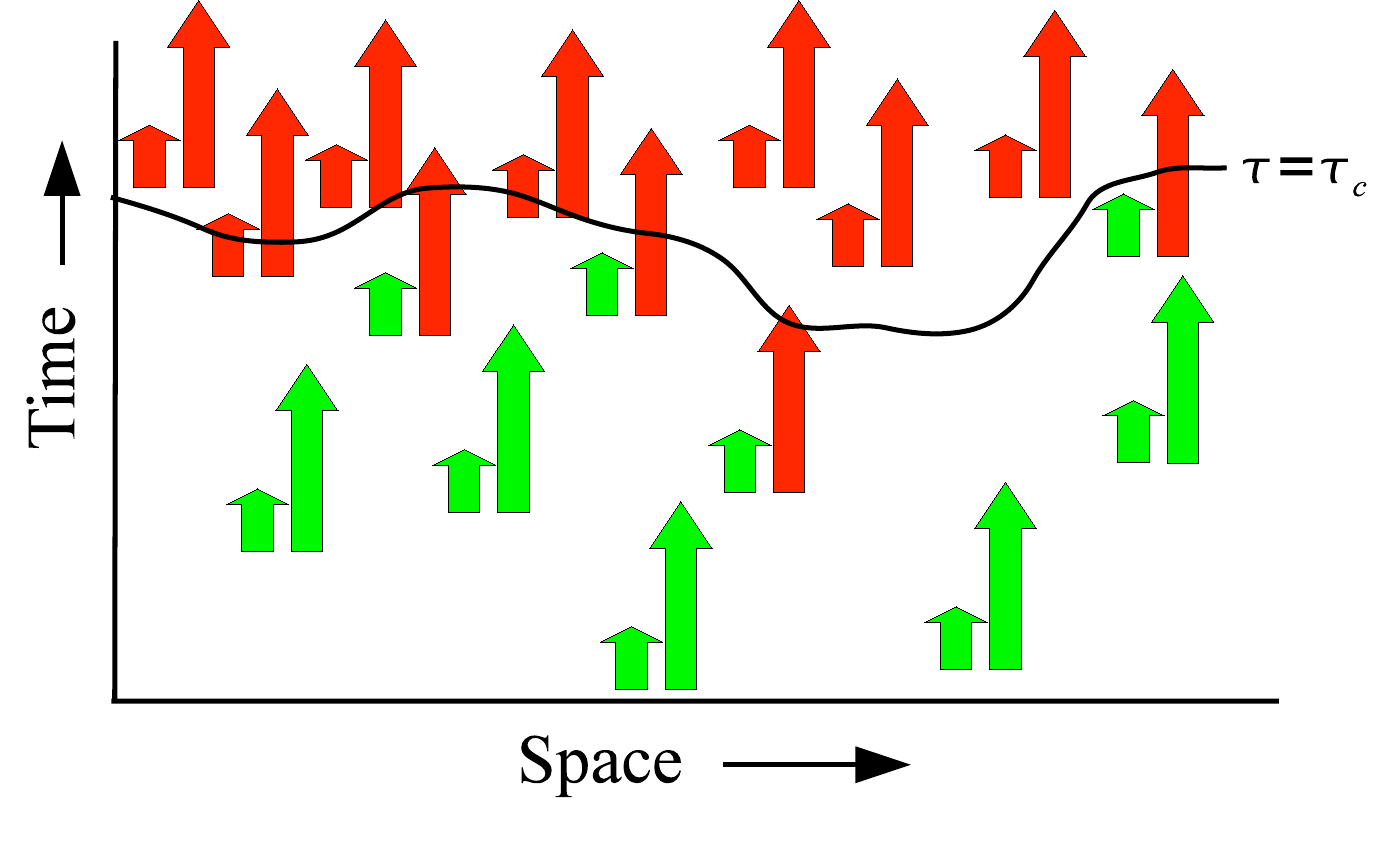}
   \caption{Counting experiments in the G-V paradox.}
   \label{fig:paradox}
   \end{center}
   \end{figure}

The thought experiment becomes a paradox when we imagine that a
subject wakes up, with no information about how long he slept,
and is asked the probability that he is the designated Head
Subject, who had slept for one minute.  The intuitive answer
would be 50\%, since it was determined by the flip of a fair
coin, but this is not what the global time cutoff measure gives
(See Fig.\ \ref{fig:paradox}%
). To apply the global time cutoff measure, one must count the
number of awakenings from one-minute naps and from one-hour naps
in the sample spacetime region (lighter, or green arrows). But
the number of one-hour naps (long arrows) is equal to the number
of experiments that started more than one hour before $\tau_c$,
while the number of one-minute naps (short arrows) is equal to
the number of experiments starting more than one minute before
$\tau_c$.  The exponential expansion implies that there are $Z$
times as many awakenings from short naps, so the probability of
being the Head Subject is not 50\%, but instead $Z/(Z+1)$, which
can be very close to 1.

Thus, according to the global time cutoff measure, the
probability of the outcome of the coin flip is no longer 50/50
when the subject awakes, but instead it has become more probable
that the outcome was the one that led to the one-minute nap.  If
we imagine $Z \gg 1$, then when the subject wakes up he can be
almost certain that he is waking from a short nap. 

To make things sound even more bizarre, suppose that when the
subject wakes up he is not told the outcome of the coin or the
length of his nap.  Suppose instead he goes back to sleep, this
time being told that he will be awakened at 60+$\epsilon$ minutes
after the original coin toss, regardless of the outcome.  When he
awakes, he is again asked what is the probability that his
original nap was the short one.  Now, according to the global
time cutoff measure, we count all events in which the subject
wakes up for the second time, which is always 60+$\epsilon$
minutes from the start.  This time there is no bias imposed by
the measure.  Even though the subject was almost sure when he
woke up the first time that he was awakening from a short nap,
now he concludes that the probability is 50\%, as he had thought
immediately after the coin flip.

These results seem strange, but BFLR point out that they are
perfectly understandable if we interpret the cutoff as a physical
end of time.  In that case, when the subject goes to sleep, he
has no guarantee that he will wake up.  Perhaps the end of time
will occur while he is asleep, and then it is all over.  Thus he
learns something clearly new when he awakes: he learns that time
has not ended yet.  This is more likely to be the case if he is
awakening from a one-minute nap than if he is awakening from an
hour nap, so the enhanced probability of the short nap is implied
by a straightforward conditional probability calculation.  If he
goes to sleep for a second time and is awakened at 60+$\epsilon$
minutes from the start, he has to recalculate the conditional
probabilities.  Now he knows that it is 60+$\epsilon$ minutes
after the coin flip and time has not ended, but the probability
for that is independent of the outcome of the coin flip.  The
conditional probability is now unbiased, so it is 50\%.

We agree that the G-V paradox can easily be understood if the
global time cutoff is taken as a physical end of time.  However,
we will argue that it can also be understood --- although in a
more subtle way --- in the mathematical limit interpretation. 
Before describing this argument, however, we will first review
some basic facts about the properties of probabilities defined on
infinite sets.

\section{Probabilities on Infinite Sets}

Suppose we consider the sequence of integers in their normal
order,
\beq
  S = (1,\ 2,\ 3,\ \dots) \ .
\label{eq:normal}
\eeq
Suppose further that we are given some bounded function $f(n)$
defined on the integers.  An example of such a function might be
the function which distinguishes the even integers,
\beq
  f(n) = \lcurl \begin{array}{ll}1 & \hbox{if $n$ is even}\\
     0 & \hbox{if $n$ is odd.}\\
           \end{array}\right.
\eeq
If we could define $\expect{f(n)}$, the average value of $f(n)$,
then we have defined the fraction of integers that are even. 
This question is famously ill-defined, so there is no unique
answer, but it is nonetheless possible to define a procedure
which makes the answer unique.  In particular, one simple choice
is what we will call the sequential cutoff measure, in analogy to
the global time cutoff measure.  Specifically, we can define
\beq
  \expect{f(n)} \equiv \lim_{N \to \infty} \, {1 \over N} \,
     \sum_{n=1}^N f(n) = \half \ .
\label{eq:seqcutoff}
\eeq
This gives a unique answer, but the uniqueness depends on
processing the integers in their standard sequential ordering,
$S$.  It is well-known, however, that one can reach a different
conclusion by ordering the integers by starting with the first
two odd integers, then listing the first even integer, then the
next two odd integers, then the next even integer, etc.:
\beq
  K = (1,\ 3,\ 2,\ 5,\ 7,\ 4,\ 9,\ 11,\ \dots) \ .
\label{eq:K_n}
\eeq
Every integer occurs once and only once on this list, so it
represents a re-ordering of the set of all integers.  If the
sequential cutoff measure is applied to this sequence, then
\beq
  \expect{f(n)} \equiv \lim_{N \to \infty} \, {1 \over N} \,
     \sum_{n=1}^N f(K_n) = {1 \over 3} \ ,
\label{eq:reordered}
\eeq
where $K_n$ denotes the $n^{\rm th}$ integer in the sequence
$K$\null. Thus, the sequential cutoff measure defines the
fraction of integers that are even for any particular ordering of
the integers, but the answer depends on the ordering.

To illustrate the relevance of the sequential cutoff measure, we
quote an important mathematical theorem, discovered by \'Emile
Borel \cite{Borel} in 1909, that makes use of it.  The theorem is
a special case of the Strong Law of Large Numbers.  Consider a
real number $r$ in the range $0 < r < 1$.  Such a number can
always be expanded as a binary fraction, which is an infinite
sequence of 0's and 1's, such as
\[
r = 0.11010001011 \ \dots \ .
\]
If we let $B_n$ denote the n'th binary digit after the point,
then the fraction $\Phi(r)$ of digits that are 1's can be defined
by the sequential cutoff measure as
\beq
  \Phi(r) \equiv \lim_{N \to \infty} \, {1 \over N} \,
     \sum_{n=1}^N B_n \ .
\label{eq:phi(r)}
\eeq
$\Phi(r)$ is a real number that ranges from 0 to 1, depending on
the argument $r$.  We might expect that values at or near 1/2
would be preferred, since there is no reason why a given digit
should be more likely to be a 1 than a 0.  Suppose we consider
the set of all real numbers $r$ between 0 and 1 for which
$\Phi(r)$ is exactly equal to 1/2.  The theorem states that the
Lebesgue measure of this set is 1.  That is, except for a set of
measure zero, every real number has exactly half 0's and half 1's
in its binary expansion, as defined by the sequential cutoff
measure of Eq.~(\ref{eq:phi(r)}).

Our basic point is that mathematicians have had over a century of
experience with what we are calling the sequential cutoff
measure, and they have not found any inconsistencies.  No one has
argued that the use of this definition requires a maximum
possible integer.  At any stage in the calculation of the limit
one is considering only a finite set of $N$ integers, with a
maximum integer $N$, but in the limit the notion of a maximum
integer disappears.  The global time cutoff measures are based on
essentially the same idea, where the sum over functions of
integers in the sequence is replaced by the sum over stories in
the multiverse, ordered by the global time of their occurrence.

BFLR argue that the cutoff in a global time cutoff measure cannot
be claimed to disappear in the limit, since a nonzero fraction of
observers survive until the cutoff, even as $\tau_c \to \infty$. 
It is true that a nonzero fraction of observers survive until
$\tau_c$ for any $\tau_c$, but we question whether this statement
implies that any observer sees the end of time. 

In considering whether the nonzero-fraction statement implies an
end of time, it is useful to consider the analogous question for
the integers, where we have more experience.  For the integers
and the sequential cutoff measure, we will find that the fraction
of integers that are within some fixed distance from the cutoff
will go to zero as $N \to \infty$, so the situation at first
looks different.  But this is not the only question that can be
asked.  Suppose we ask what fraction of the integers are so large
that they cannot be doubled without exceeding the cutoff $N$. 
That fraction will approach 1/2 in the limit as $N \to \infty$. 
Does this statement imply that there is an end to the integers,
and that half of the integers are so large that they cannot be
doubled? Of course not, since we know from the Peano axioms that
there is no end to the integers.  The statement merely implies
that as the limit is taken in evaluating
Eq.~(\ref{eq:seqcutoff}), at any stage half of the integers
included in the sum are larger than $N/2$.  But $N$ is not the
largest integer; it is simply a variable that is introduced to
define a limiting process.  Once the limit is taken, there is no
integer which is so large that it cannot be doubled.  Similarly,
as the limit $\tau_c \to \infty$ is taken in
Eq.~(\ref{eq:events}), a fixed fraction of all births included in
the sum will occur at a time later than $\tau_c - \Delta \tau$,
where $\Delta \tau$ is an observer lifetime.  But, like the
variable $N$ in Eq.~(\ref{eq:seqcutoff}), $\tau_c$ in
Eq.~(\ref{eq:events}) is simply a variable that is introduced to
define a limiting process.  Once the limit is taken, there will
be no observers who reach the end of time, just as there are no
integers so large that they cannot be doubled.  An end of time
will occur only if the multiverse model itself has been modified
to end at some specific time.

\section{Resolution of the Guth-Vanchurin Paradox}

\subsection{Step by step construction}
\label{sec:stepbystep}

Now we can describe our understanding of the Guth-Vanchurin
paradox, in the context of a regularization procedure that
manifestly does not incorporate an end of time.  We believe that
the best way to understand the thought experiment is to build it
up in stages, one stage at a time.

As a first stage, let us consider the experiment with the
subjects left out.  There is only the experimenter.  Since the
flip of the coin was used only to assign head and tail
designations to the subjects, we can also leave out the coin.  We
will assume, however, that the experimenter still sets two alarm
clocks, one for one minute, and one for one hour.  She also
writes, but does not submit, two reports.  The first, which we
call the head report, announces that it is time for the Head
Subject to wake up.  The second, the tail report, announces the
awakening of the Tail Subject.  The subjects are not present, but
when the first alarm clock goes off, the experimenter sends the
head report to the arXiv.  When the second alarm clock rings, she
sends the tail report.  Each experiment results in one head
report and one tail report, but the head report is submitted 59
minutes earlier.

At this stage there are no random elements in the description,
and no need to discuss a probability measure, but there is
already a peculiar consequence.  If we imagine that experiments
of this type are occurring throughout the multiverse, then the
total rate of arrival of head reports at all the arXivs of the
multiverse will be equal to the rate of experiments that began
one minute earlier, while the rate of arrival of tail reports
will be equal to the rate of experiments that started one hour
earlier.  The rate of arrival of head reports is therefore larger
by the factor $Z$, the expansion factor of the multiverse
corresponding to 59 minutes:
\bea
&&\hbox{Rate of arrival of head reports} =\nonumber \\ &&\qquad Z
\times \hbox{Rate of arrival of tail reports.}
\label{eq:arrivals}
\eea

The result here may seem counter-intuitive, but it is purely a
matter of counting, and it is an unavoidable feature of
exponentially expanding systems.  The rate of observation of a
given outcome can be biased by a time delay in the reporting. 
Even though the head and tail reports are created in matched
pairs, the head reports arrive at the arXiv at a rate that is $Z$
times larger than the rate for tail reports.  This result is
independent of the probability measure, although it does depend
on the choice of a global time variable $\tau$ which is used both
to measure $Z$, and to clock the rate of arrivals.  This time
delay bias is of course just an example of the youngness bias
discussed in Sec.~\ref{sec:youngness}.

As a second stage, we can introduce a probability element by
considering a fictional multiversal arXivist, who records the
arrival of all the reports to all the arXivs in the multiverse
(or at least all the reports in the sample spacetime region of
the mathematical model of the multiverse).  When a report is
posted, the arXivist learns about it immediately, at the same
value of the global time coordinate $\tau$.  (The reader may be
shocked by our gross disregard for the limits imposed by the
speed of light, but remember that we are discussing the
consistency of a measure that from the beginning has been based
on the statistics of counting all events up to a final cutoff
time $\tau_c$.  Such calculations can in principle be done in the
mathematical model of the multiverse, even if they would be
physically impossible in the real multiverse.  In discussing the
consistency of the probabilities defined in such a global time
cutoff measure, it makes sense to consider hypothetical observers
who have access to the same information that is being used to
define these probabilities.) Suppose the arXivist is asked to
determine the odds that a given report, chosen randomly from the
stream of reports arriving at the arXiv, is a head report.  Since
the arXivist knows that the report was chosen randomly from a
stream in which the relative rates of arrival of head and tail
reports is given by Eq.~(\ref{eq:arrivals}), he would conclude
that a head report is $Z$ times more likely than a tail report. 
The probability is well-defined despite the fact that he will see
an infinite number of reports over time, because the ordering in
which he sees them is fixed.  The probability that the $N^{\rm
th}$ report he receives is a head, for any $N$, is a well-defined
calculation for which the answer approaches $Z/(Z+1)$ for large
$N$.  The arXivist would attribute the asymmetry between head and
tail reports to the observational bias described in the previous
paragraph.  Since heads and tails are reported with different
time delays, the one with the shorter time delay is observed with
higher probability.  The deviation of the probability from 50\%
has nothing to do with anybody falling asleep or waking up, but
is simply a consequence of the experimental protocol, which
prescribes different time delays for head and tail reports.

Before we go on, let us look a bit more into the question of how
probabilities can change.  Immediately after the reports were
written, a random report would have an equal chance of being a
head or a tail.  How can it be, then, that the probability is
$Z:1$ in favor of head reports when they arrive at the arXiv? The
answer, we would argue, is exactly the same as the difference in
the expectation values calculated in Eqs.~(\ref{eq:seqcutoff})
and (\ref{eq:reordered}).  The statistics of infinite sets can
depend on their ordering.

We can make the correspondence explicit by imagining that all the
reports written in the multiverse are numbered according to the
global time at which they are written, with the head report being
numbered by convention ahead of the tail report that is written
at the same time.  The list of all reports, each described by its
number with a subscript to indicate heads or tails, would then
look like:
\beq
  1_H,\ 2_T,\ 3_H,\ 4_T,\ 5_H,\ \dots \ .
\label{eq:normalHT}
\eeq
However, since the head reports reach the arXiv faster, the
listing in the order of arrival might look instead like
\beq
  1_H,\ 3_H,\ 2_T,\ 5_H,\ 7_H,\ 4_T,\ 9_H,\ 11_H,\ \dots
     \ .
\label{eq:reorderedHT}
\eeq
The change in the fraction of reports that are head reports is,
in this construction, identical to the change in the fraction of
integers that are odd, as the integers are re-ordered from
Eq.~(\ref{eq:normal}) to Eq.~(\ref{eq:K_n}). 

BFLR discuss the probabilities of short naps, in the G-V thought
experiment, in the following terms: ``\dots \ there are two
reference classes one could consider. When going to sleep we
could consider all people falling asleep; 50\% of these people
have alarm clocks set to wake them up after a short time. Upon
waking we could consider the class of all people waking up; most
of these people slept for a short time. These reference classes
can only be inequivalent if some members of one class are not
part of the other. This is the case if one admits that some
people who fall asleep never wake up, but not if one insists that
time cannot end.''

The logic used by BFLR would certainly be compelling if we were
talking about finite reference classes.  However, when infinite
classes are involved, the logic no longer holds.  The lists in
Eqs.~(\ref{eq:normalHT}) and (\ref{eq:reorderedHT}) both describe
the same set, the set of all integers, and hence the same
reference class.  There are no members of one class that are not
part of the other.  Yet the second list looks like 2/3 of its
elements are heads, a conclusion that can be made precise by the
sequential cutoff measure.

To complete the buildup of the G-V thought experiment, we can add
in the subjects.  Now the alarm clocks both wake their
corresponding subjects, and trigger the sending of their
corresponding reports.  We can describe this compactly by
imagining that the head and tail reports are each written to
CD's, with the head report placed under the pillow of the Head
Subject, and the tail report under the pillow of the Tail
Subject.  Upon being awakened by the alarm clock, the subjects
have been instructed to put the CD into their computer and send
the report to the arXiv. 

In addition, the subjects upon awakening are asked the
probability that they had a short nap, or equivalently the
probability that the report on their CD is a head report.  But
the knowledge that the subject has, in this situation, is
essentially identical to the knowledge that the arXivist has when
he receives the report.  They both know the experimental
protocol, but they know nothing about the particular report that
would make it any different from any other report.  The arXivist
concluded unambiguously that the probability of a head report is
$Z/(Z+1)$, but the situation for the subject is not so clear. 
The arXivist knew that he would receive all reports and he knew
in what order he would receive them, but the subject has only one
CD and must choose a measure in order to determine a probability. 
If the subject chooses to use the global time cutoff measure
based on the global time variable $\tau$, then he will calculate
the total number of head and tail reports arriving at all the
arXivs in the multiverse before some cutoff time $\tau_c$, and he
will reach the same conclusion as the arXivist: the odds are
$Z:1$ in favor of the shorter nap.  We cannot argue that this is
what the subject will do, or that it is what he should do.  We
are not trying to argue that any particular measure is correct. 
However, if he decides to use this method to determine
probabilities, then we should not have any trouble understanding
his logic.  He is simply taking the statement about frequencies
in Eq.~(\ref{eq:arrivals}), and using this to infer a
probability.  As was the case for the arXivist, the deviation of
the probability from 50\% has nothing to do with anybody falling
asleep or waking up, but is simply a consequence of the
experimental protocol, which prescribes different time delays for
head and tail reports.  Provided that the subject has chosen to
use this measure, he would have concluded as soon as he learned
the experimental protocol that the odds would be $Z:1$ in favor
of the shorter nap.

Finally, let us consider what happens if the subject is not told
the outcome when he wakes, but instead is put back to sleep until
60+$\epsilon$ minutes from the original coin toss.  During the
brief period of wakefulness before he goes back to sleep, the
subject (assuming the global time cutoff measure) judges the odds
to be $Z:1$ in favor of being the Head Subject, with the short
nap.  But now the experiment is going to be done in reverse.  If
he is the Head Subject, he will be going to sleep for a 59-minute
nap.  If he is the Tail Subject, it will be a nap of length
$\epsilon$ minutes. Since the probability of observing a
particular outcome is biased by the difference in time delays,
the second nap --- or time delay --- will bias the probability in
favor of being the Tail Subject.  And since the time difference
is again 59 minutes, the bias will again be a factor of $Z$. 
This extra bias will turn the $Z:1$ odds back to $1:1$, which
agrees with the conclusion in Sec.~\ref{sec:Guth-Vanchurin}.

\subsection{Betting on the experiment as a test of consistency}

It is certainly possible to talk about probabilities without
betting, and physicists and mathematicians do that all the time. 
Nonetheless, it is sometimes useful to think about betting as a
way of clarifying and double-checking our thoughts about
probabilities.  In particular, the G-V thought experiment
introduces a situation where the time at which a hypothetical bet
is to be paid can depend on the outcome, and that introduces an
important issue that we have not yet discussed.

Our goal in discussing betting is simply to make sure that the
probabilities are understood correctly.  With that in mind, we
should be aware that any real bettor will have personal
preferences that will affect his betting.  Some bettors might
prefer the thrill of betting on long-shots, where the probability
of winning is low but the payoff is large.  Others might hate
such bets.  If there is a situation where alternative outcomes
might lead to payoffs at different times, different bettors might
have their own preferences about whether it is better to receive
\$$X$ now or \$$Y$ tomorrow.  One way of dealing with these
differences would be to describe a precise model of the
preferences of the bettor to be discussed.  We, however, will
avoid these issues by focussing only on determining the betting
odds for which the bettor breaks even.  That is, for what odds is
the expectation value of the bettor's winnings equal to zero?

Since we are discussing events in an infinite multiverse, these
expectation values are ambiguous until one chooses a probability
measure.  We should expect consistency only if we use the same
measure throughout, which is what we will do.  Once again, we are
only trying to show that the global time cutoff measures are
consistent, not that they are necessarily correct.

Consider first what would happen if the subject were asked to
make an immediate bet on whether he is assigned a head or a tail,
at the time of the coin flip.  By counting stories under the
cutoff one would find equal numbers of head and tail assignments,
so the break-even bet is 50/50.  If the subject bets with even
odds, he will break even.  Note that if we are examining the
break-even point, there is no loss in generality by assuming that
the subject always bets on heads.

Now consider a second type of bet, in which the subject is asked
to bet when he wakes up, with the understanding that the payoff
would occur immediately after the bet is placed.  Then, as
discussed earlier, the subject would see the odds as $Z$ to 1 in
favor of heads, so his break-even bet would be to use these odds. 
That is, he would break even if he bets that he is paid one
dollar if he is found to be the Head Subject, and he pays $Z$
dollars if he turns out to be the Tail Subject.

To see that this bet implies no net gain or loss in the
multiversal expectation value, imagine a multiversal statistician
keeping track of all the events in the sample spacetime region,
which ends at $\tau = \tau_c$.  Every time a paper is posted to
the arXiv, an experiment is completed and money changes hands. 
But the arrival rate of head reports will exceed that for tail
reports by a factor of $Z$, which is exactly what is needed for
the subjects to break even on their bets.

Note, by the way, that it does not matter whether the subject is
asked to choose his bet before or after he goes to sleep.  As
soon as the nature of the experiment and the terms of the bet are
described, the subject will conclude that this is a break-even
bet. 

Now consider a third type of bet, in which the payoff occurs at
60+$\epsilon$ minutes after the subject has gone to sleep.  Thus,
at the time of the payoff, the subject will have just woken up if
he is the Tail Subject, and would have woken up 59 minutes
earlier if he is the Head Subject.  By counting events of each
type in the sample spacetime region, we see that there are equal
numbers of each outcome.  The break-even betting odds are then
even.  Of course, to have a fair bet it is necessary that the
bettor does not know the outcome when he places the bet.  Thus,
for this to be a fair bet, we might imagine that the subject is
asked to place his bet before he goes to sleep.  Alternatively,
we might imagine that the subject is put back to sleep
immediately after waking up the first time, as described in the
last paragraph of the previous subsection.  Either of these
choices will allow a fair bet, and the subject will break even if
he bets with even odds.

But now one might worry about the consistency of the bets of the
second and third types: if the subject is almost sure that he is
associated with heads when he wakes up, how can he think the odds
are 50/50 at one hour after the experiment started?

\begin{table}[t!]
\includegraphics[width=5.42in]{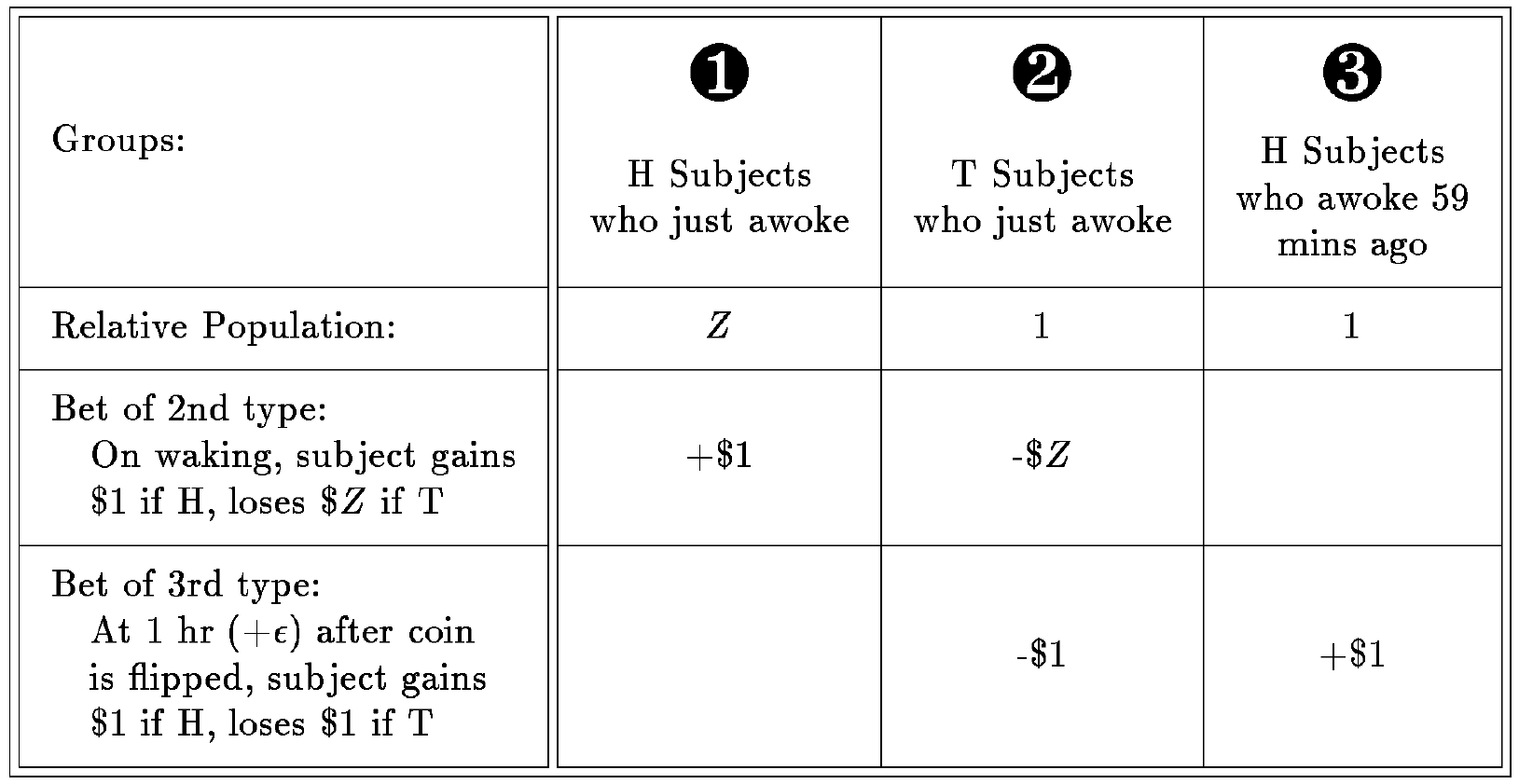}
\caption{Three groups that need to be counted in the sample
     spacetime region to determine the expectation value of the
     results of two kinds of bets.  The bets are discussed in the
     text, and summarized in column 1.}
\label{tab:gvtable}
\end{table}

It is fairly straightforward to understand these results from the
point of view of a multiversal statistician, who is counting all
the events that happen in the sample spacetime region, which ends
at the cutoff.  Let us try to simultaneously consider bets of the
second and third types, as listed in Table~\ref{tab:gvtable}. 
For the second type, the payoff is at the time of waking up, and
the subject gains \$1 if he turns out to be the Head Subject, and
loses $Z$ dollars if he turns out to be the Tail Subject.  For
the third type, the payoff is 60+$\epsilon$ minutes after the
coin was flipped and the subjects went to sleep, and the subject
gains \$1 if he turns out to be the Head Subject, and loses \$1
if he turns out to be the Tail Subject.  At any given time, there
are three groups of subjects that the multiversal statistician
should want to keep track of: (1) the Head Subjects who just woke
up; (2) the Tail Subjects who just woke up; and (3) the Head
Subjects who woke up 59 minutes earlier. Group (1) is bigger than
group (2) by a factor of $Z$, but group (3) is the same size as
group (2).  For bets of the second type, the subjects in group
(1) each gain \$1, while the subjects in group (2) each lose $Z$
dollars.  Thus the expectation value for the subject is to break
even.  For bets of the third type, subjects in group (2) each
lose \$1, and subjects in group (3) each gain \$1, and again the
expectation value is to break even.  This pattern is stable over
time; 59 minutes later the subjects of group (1) become the
subjects of group (3) for the later time.  They are just as
numerous at the later time as they were at the earlier time, but
at the later time they are outnumbered by the subjects in the new
group (1) --- the subjects associated with heads who are waking
up at the later time.  These subjects outnumber their
counterparts from the earlier time by a factor of $Z$.

From the point of view of the subject, however, these results may
seem rather surprising.  But if the subject chooses to use the
global time cutoff to determine his probabilities, he would do
the same calculations as those of the multiversal statistician
described above, and would reach the same conclusion.  He would
describe the change in the probability, between the bet of the
2nd type and the bet of the third type, as the result of the time
delay bias that we discussed in the previous subsection.  Whether
the subject is required to declare his bet before he goes to
sleep, or whether he is put back to sleep after waking up the
first time, there is a differential time delay between the two
bets.  If he is the Head Subject, the third type of bet occurs 59
minutes after the 2nd type of bet.  If he is the Tail Subject,
there is no delay.  In global time cutoff measure, whenever there
is a different time delay for different experimental results, the
observation of those results is biased by the time delay. If the
subject has become accustomed to the time-delay bias, he would
understand how bets of the 2nd type are consistent with bets of
the third type.

\subsection{What if the time of the payoff of a bet depends on
the outcome?}

We consider now the point of view of the experimenter.  She is
awake the whole time, so she always knows what time it is, and
can observe the subjects whether they are asleep or awake.  At
one minute after the coin flip, she observes with certainty that
the Head Subject wakes up, while the Tail Subject continues to
sleep.  If the subjects' real names are Rosencrantz and
Guildenstern, then the experimenter would calculate, by counting
stories in the sample spacetime region, that there is a 50\%
chance that Rosencrantz will be designated the Head Subject at
the time of the coin flip, and also a 50\% chance that
Guildenstern will.  At any other time, she would also calculate
that each has a 50\% chance of being the Head Subject.  At one
minute after the coin flip, she would find that Rosencrantz has a
50\% chance of waking up, and a 50\% chance of continuing to
sleep.  At one hour she would see with certainty that the Tail
Subject wakes up, and with 50\% probability it would be
Rosencrantz, precisely in those cases in which Rosencrantz did
not wake up at one minute.  In short, none of these probabilities
have been influenced by the multiverse.  The experimenter can
publish her results, but they will show no evidence for the
multiverse or the measure.  The enhanced probability of the short
nap is relevant only to the subjects.

Suppose, then, that Rosencrantz offers to bet with the
experimenter, betting that he will be the Tail Subject.  The bet
is to be paid off when Rosencrantz wakes up.  Knowing the $Z:1$
odds according to the global time cutoff measure, Rosencrantz
proposes that he would pay the experimenter \$1 if he turns out
to be the Head Subject, but the experimenter should pay him \$$Z$
if he turns out to be the Tail Subject.  From Rosencrantz's point
of view, this is certainly a break-even bet, given the global
time cutoff measure.  But how does the experimenter view it?

If it is a break-even bet for Rosencrantz, it must be a
break-even bet for the experimenter, too, as calculated by global
time cutoff measure expectation values.  The expectation value is
calculated, after all, by simply counting transactions in the
sample spacetime region.  In each transaction, any gain for
Rosencrantz is a loss for the experimenter, and vice versa.  If
the transactions average to zero for Rosencrantz, then of course
they must also average to zero for the experimenter.  But to the
experimenter, there is a 50\% chance that Rosencrantz will be the
Head Subject, in which case the experimenter will receive \$1 at
1 minute after the coin flip.  There is a 50\% chance that
Rosencrantz will be the Tail Subject, in which case the
experimenter will have to pay \$$Z$ at 1 hour after the
coin flip.  How is this breaking even?

If one looks at the Rosencrantz/experimenter transactions in the
sample spacetime region, one sees that there are $Z$ times as
many \$1 payments from Rosencrantz to experimenter as there are
\$$Z$ payments from experimenter to Rosencrantz; the former
happen at a time that is earlier by 59 minutes, so $Z$ times as
many fit under the cutoff.  Thus, as an expectation value
calculated in the global time cutoff measure, the experimenter
does break even.  The experimenter has a 50\% chance of winning
\$1 at 1 minute, and a 50\% chance of losing \$$Z$ at 1 hour, but
in the youth-biased multiverse there are $Z$ times more
experimenters at 1 minute as there are at 1 hour.

From the experimenter's own perspective, having the opportunity
to win \$1 at 1 minute or lose \$$Z$ at 1 hour, each with 50\%
probability, may or may not sound attractive.  She may have her
own preferences about how desirable it is to have money sooner
rather than later.  If, however, there is a freely available
banking system that lets her exchange money at some fixed
interest rate, then she can evaluate this bet independently of
her personal preferences for having money sooner vs.\ later. 
Suppose, for example, that accrued interest for 59 minutes
increases the value of a deposit by a factor $\bar Z$.  Then the
experimenter can settle the bet at 1 minute, if she chooses, by
collecting the \$1 if she wins, or depositing \$$(Z/\bar Z)$ if
she loses, so the bank account could pay the debt at the end of
the hour.  Alternatively, if she prefers to settle the bet at 1
hour, she can deposit the \$1 at one minute if she wins, and then
at 1 hour she would either obtain \$$\bar Z$ from the bank, or
pay \$$Z$ to Rosencrantz.  Thus, for either of these choices, she
would view this as breaking even if $\bar Z = Z$.  This can be
seen to be a general theorem:

\begin{itemize}
\item[] {\it Multi-Payoff-Time Betting Theorem:}
If a bet involves payoffs at a time that depends on the outcome,
the break-even point for the global time cutoff measure can be
found by assuming that the interest rate is equal to the
multiversal expansion rate.
\end{itemize}

The above theorem serves to clarify an important point concerning
experiments and the probabilities for their outcomes.  When we
specified the probability of the outcome of an experiment in
Eq.~(\ref{eq:expt}), we did not require that the different
outcomes all end at the same time.  The probabilities are defined
by counting the number of stories under the cutoff, so stories
that end earlier are favored.  Suppose that we wish to bet on
such an experiment, and we want to understand what the
probabilities of Eq.~(\ref{eq:expt}) are telling us about how to
find the break-even odds.  To use these probabilities directly,
we have to assume that the payoff that would result for a
particular outcome $S_i$ would occur exactly when $S_i$ ends, so
that the frequency of the payoff (counted in the sample spacetime
region) would match the frequency of the experimental result.  If
the payoff came earlier or later, its frequency in the sample
spacetime region would be different.

\begin{figure}[htbp]
   \begin{center}
   \includegraphics{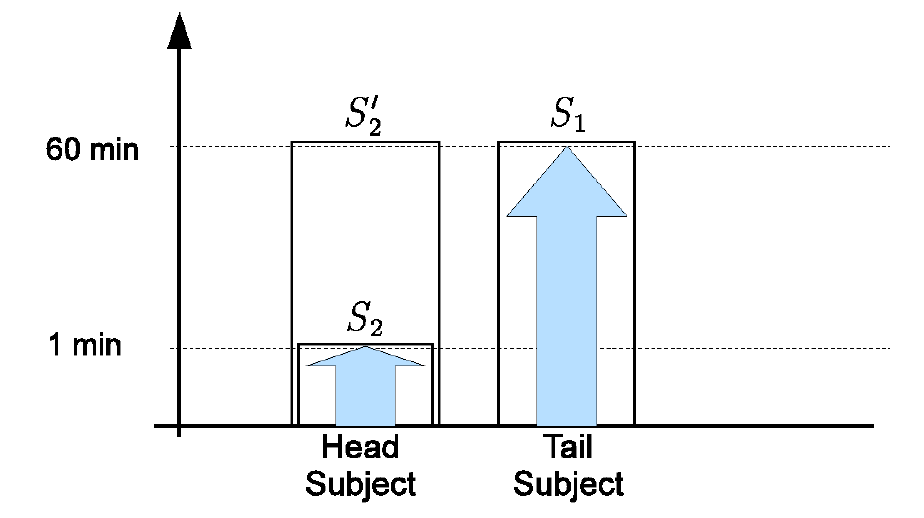}
   \caption{G-V experiment viewed with two different ending points.}
   \label{fig:expt}
   \end{center}
   \end{figure}

So, when the experimenter thinks about his prospective bet with
Rosencrantz, he can view the experiment in either of two ways, as
shown in Fig.\ \ref{fig:expt}.
We will call the losing outcome $S_1$, where Rosencrantz wakes up
after 1 hour, terminating the experiment.  But the winning
outcome could reasonably be described in either of two ways.  In
the first, which we will call $S_2$, Rosencrantz wakes up at 1
minute, and the experiment ends.  In the second, which we will
call $S_2'$, Rosencrantz also wakes up at one minute, but in this
version the experiment is not considered over until one hour from
the coin flip.  Thus, $S_2$ ends at 1 minute, while $S_1$ and
$S_2'$ end at one hour.  Since an experiment is described in
Eq.~(\ref{eq:union}) as a union of outcomes, the experiment can
be described as either $S_{\rm expt} = S_1 \cup S_2$, or as
$S_{\rm expt}' = S_1 \cup S_2'$.  For $S_{\rm expt}$ the
probabilities are $Z:1$ in favor of outcome 2, while for $S_{\rm
expt}'$ the probabilities are 50/50.  Both of these answers are
right, but they are the answers to different questions.  If the
payoffs are to be made when the subject wakes up, then experiment
$S_{\rm expt}$ is the right description, and the break-even bet
is based on $Z:1$ odds.  If, however, the payoff is made at one
hour after the coin flip, regardless of outcome, then $S_{\rm
expt}'$ is the appropriate description, and the break-even bet is
50/50.  Both Rosencrantz and the experimenter will agree that
these are the break-even bets in both cases, according to the
global time cutoff measure, and the two bets can be seen to be
equivalent to each other by assuming that the interest rate is
equal to the multiversal expansion rate.

The relevance of an interest rate equal to the multiversal
expansion rate is perhaps made clearer by considering a fictional
multiversal bank.  The multiversal bank will freely lend money or
accept deposits with an interest rate equal to the multiversal
expansion rate.  All transactions are assumed to be instantaneous
in the global time coordinate $\tau$ that is used to define the
cutoff.  While such a bank is clearly impossible, it is
nonetheless interesting that, if such a bank did exist, it would
break even according to the global time cutoff measure. 
Consider, for example, one-year certificates of deposit.  On any
given day, deposits of this type might total \$X.  One year later
these depositors will come back and collect $\tilde Z \times
\$X$, where $\tilde Z$ is the expansion factor of the multiverse
over one year, $\tilde Z = \exp(\lambda_c \times 1 \hbox{ yr})$.
But the universal exponential growth of the multiverse ensures
that on the same day, deposits to the one-year certificate
accounts will be $\tilde Z$ times larger than the previous year,
and hence $\tilde Z \times \$X$, so the bank breaks even.  On
Earth this would be called a Ponzi scheme, but it is sound
practice for a fictional multiversal bank.  Money lending with an
interest rate equal to the expansion rate will always break even
in the global time cutoff measure.

Finally, we point out that the use of the global time cutoff
measure does not in any way restrict the betting preferences that
a person could choose.  In the G-V thought experiment, while
everyone should agree that the global time cutoff measure implies
that the break-even odds for the experiment $S_{\rm expt}$ are
$Z:1$, everyone is also allowed to have his own betting
preferences.  Betting preferences and the calculation of
probabilities are two independent issues.  For example,
Rosencrantz might be convinced that the global cutoff measure is
correct, but his primary concern might not be the expectation
value for the money exchange immediately after he wakes up.  He
might instead be saving to pay his rent, which might be due 60
minutes after the start of the experiment.  Since he is
interested in making predictions for the time when his rent is
due, he would calculate the probabilities for experiment $S_{\rm
expt}' = S_1 \cup S_2'$, which ends at the relevant time.  The
probability is then 50/50.  Since the payoff is nonetheless
scheduled to take place at 1 minute if he is the Head Subject, he
will need to use an interest rate to convert this probability
into a statement about break-even odds.  If he knows that the
actual interest rate available to him would increase the value of
a deposit by a factor \$${\bar Z}$ over 59 minutes, he would
conclude that to break even by his personally preferred
criterion, the odds should be $\bar Z:1$ in favor of heads.  For
the special case $\bar Z = Z$ this would be identical to
experiment $S_{\rm expt} = S_1 \cup S_2$, but otherwise they are
different.  But for any preference and for any available interest
rate, the global time cutoff measure can be used to calculate the
relevant expectation values so that Rosencrantz can decide how to
bet.
For the example discussed in this paragraph, the
result is the same as what we would calculate without knowing
anything about the multiverse.

\section{Can I choose the Higgs mass?}
\label{sec:choose}

Since the G-V thought experiment leads to $Z:1$ odds in favor of
the shorter nap, is it possible to use this effect to select
where we would like to live in the multiverse?  

\begin{figure}[htbp]
   \begin{center}
   \includegraphics[width=3in]{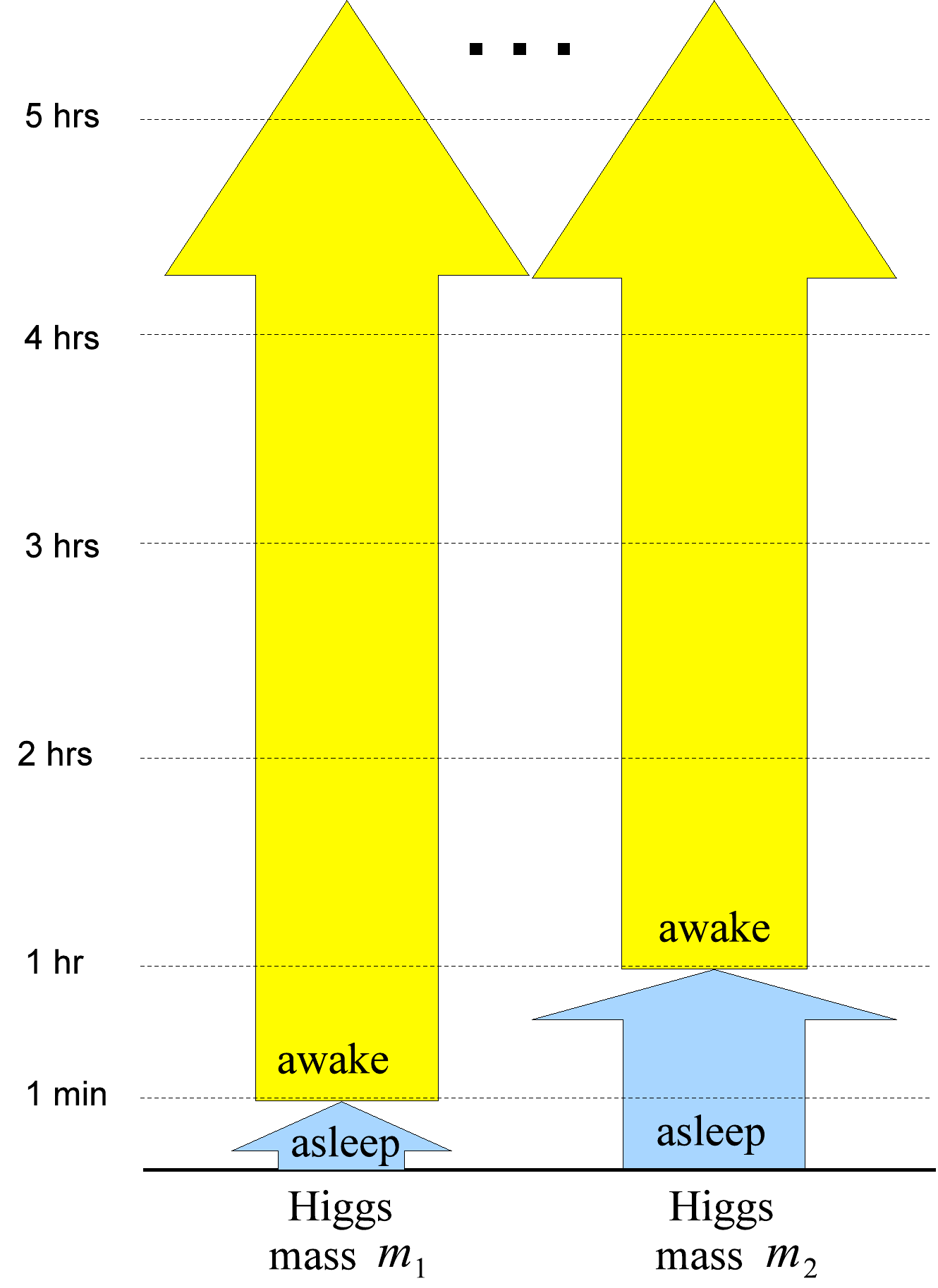}
   \caption{G-V experiment correlated with Higgs mass.}
   \label{fig:destiny}
   \end{center}
   \end{figure}

Suppose, for example, that pocket universes compatible with what
we have so far observed fall into two classes, one with Higgs
mass $m_H = m_1$, and one with $m_H = m_2$. Suppose that the full
theory is understood, and that global time cutoff calculations
(which we trust) show that we have a 50\% chance of finding
either Higgs mass.  Suppose, however, that for some reason I
strongly prefer $m_1$.  As the crucial experiment to measure the
Higgs mass is about to be done, I might consider going to sleep,
leaving instructions with a friend to wake me one minute after
the measurement if it finds $m_1$, but one hour after the
measurement if it finds $m_2$.  (As in
Sec.~\ref{sec:Guth-Vanchurin}, ``one minute'' and ``one hour''
are really stand-ins for some $\Delta t_{\rm short}$ and $\Delta
t_{\rm long}$, where the multiverse expands by a factor $Z$
between the two times.) The two possible worldlines associated
with this experiment are shown in Fig.\ \ref{fig:destiny}.  Given
what we have concluded about the G-V thought experiment, when I
wake up I should expect $Z:1$ odds in favor of Higgs mass $m_1$. 
Can I really choose where I will find myself in the multiverse?

To address this question, it is useful to be able to discuss the
probabilities as a function of time.  To accomplish this, within
the context of probabilities defined by Eq.~(\ref{eq:expt}), we
can define a generalization of the G-V thought experiment in
which the ending times are taken as variables.  That is, we can
define the outcome $S_1(t_1)$ as the story in which the subject
wakes up and finds the Higgs mass to be $m_1$, with the story
ending at time $t_1$, measured from the time when the Higgs mass
is measured.  Similarly, the alternative outcome $S_2(t_2)$ is
the story in which the subject wakes up and finds the Higgs mass
to be $m_2$, with the story ending at time $t_2$.  We then
consider the experiment $S(t_1,t_2) = S_1(t_1) \cup S_2(t_2)$. 
We further assume that the time variable $t$ used to describe the
experiment is related to the global cutoff time variable $\tau$
in such a way that ${\rm d} \tau / {\rm d} t \approx
\hbox{const}$ for the full duration of the experiment.  Since
experiments of duration $t$ must begin at a time $t$ below the
cutoff or earlier, the number of experiments of duration $t$ in
the sample spacetime region will be proportional to $e^{- \lambda
t}$, where $\lambda = ({\rm d} \tau / {\rm d} t) \lambda_c$.
Finally, we will assume for simplicity that my life span after
waking up will be exactly $t_{\rm max}$, regardless of the
outcome.  The probabilities $P(1)$ and $P(2) = 1 - P(1)$ for the
general case of experiment $S(t_1,t_2)$ determine the odds of the
break-even bet if the payoff occurs at time $t_1$ for outcome 1,
and at time $t_2$ for outcome 2.  Applying Eq.~(\ref{eq:expt}) to
this case, one finds
\beq
  P(1;t_1,t_2) = {\theta_1(t_1) e^{- \lambda t_1} \over
     \theta_1(t_1) e^{- \lambda t_1} + \theta_2(t_2) e^{- \lambda
     t_2} } \ ,
  \label{eq:P(1)}
\eeq
where $\theta_1(t_1)$ and $\theta_2(t_2)$ are given by
\bea
   \theta_1(t_1) &=& \lcurl \begin{array}{ll}
         1 & \hbox{if\ } \Delta t_{\rm short} \le t_1 \le t_{\rm
              max} + \Delta t_{\rm short}\\
         0 & \hbox{otherwise} \ ,
         \end{array}\right.
   \label{eq:theta1}\\
   \theta_2(t_2) &=& \lcurl \begin{array}{ll}
         1 & \hbox{if\ } \Delta t_{\rm long} \le t_2 \le t_{\rm
              max} + \Delta t_{\rm long}\\
         0 & \hbox{otherwise} \ .
         \end{array}\right.
   \label{eq:theta2}
\eea
Note that $\theta_1(t_1)$ is equal to 1 or 0 depending on whether
the value of $t_1$ is allowed, with a similar description for
$\theta_2(t_2)$. With the general case given by
Eq.~(\ref{eq:P(1)}), we can now consider special cases of
interest. If the experiments end at the time of waking up, $t_1 =
\Delta t_{\rm short}$ and $t_2 = \Delta t_{\rm long}$, then we
recover our previous result,
\beq
   P(1;\Delta t_{\rm short},\Delta t_{\rm long}) = {Z \over Z +
     1} \ ,
\label{eq:wakeup}
\eeq
where as before $Z = \exp \left\{\lambda \left(\Delta t_{\rm
long} - \Delta t_{\rm short}\right)\right\}$.  More generally,
this result applies whenever the two times differ by ``59
minutes,'' or $\Delta t_{\rm long} - \Delta t_{\rm short}$:
\beq
   P(1;t + \Delta t_{\rm short} ,t + \Delta t_{\rm long}) = {Z
     \over Z + 1} \ ,
\label{eq:wakeup+t}
\eeq
for any $t$ such that $0 \le t \le t_{\rm max}$.  Thus, as long
as I choose to compare the two options at equal times since
waking up, I will conclude that the Higgs mass $m_1$ is more
likely.
This result makes it seem like something has happened that is
magical: by choosing to do an experiment involving nothing more
than sleeping, I seem to have influenced the mass of the Higgs
particle.  But if we look at it more closely, the magic will
disappear.  Instead of comparing the two options at equal times
since waking up, we can compare the two options at equal clock
times $t$, measured from the instant when the Higgs mass was
measured.  In that case, we see that
\beq
   P(1;t,t) = \lcurl \begin{array}{cl}  
          1                & \hbox{if\ } \Delta t_{\rm short} < t
                             < \Delta t_{\rm long}\\
          1/2              & \hbox{if\ } \Delta t_{\rm long} < t
                             < t_{\rm max} + \Delta t_{\rm short}\\
          0                & \hbox{if\ } t_{\rm max} + \Delta t_{\rm short}
                             < t < t_{\rm max} + \Delta t_{\rm long}\ .\\
          \end{array}\right.
\label{eq:equalt}
\eeq
The expansion of the multiverse has no effect on this
calculation, since the experiment ends at the same time for
either outcome.  The Higgs mass $m_1$ is favored, but only during
the first ``hour,'' i.e., for $t < \Delta t_{\rm long}$.  I can
affect the amount of time I am likely to spend awake in a world
with $m_H=m_2$, during the first hour, by choosing to sleep
through the entire hour if the Higgs mass is equal to $m_2$. 
This is a real effect, but there is clearly nothing magical about
it.  However, whether I participate in the sleeping experiment or
not, I will find that when the clock strikes 2 hours, 3 hours, or
any later time, I am equally likely to find myself in a world
with either value of the Higgs mass.  (Under our assumptions
there is also a small difference at the end of my life, when I
will live an additional 59 minutes if the Higgs mass is $m_2$.)
While Eq.~(\ref{eq:equalt}) shows no sign of magic, it is hard to
see how it can be consistent with Eqs.~(\ref{eq:wakeup}) or
(\ref{eq:wakeup+t}).  If the probability of my living in a world
with $m_H=m_1$ is altered by my sleeping only during the first
hour and maybe the last hour of my life, how can there be another
method of accounting which says that the odds are $Z:1$ in favor
of $m_H=m_1$, where $Z$ could in principle be large? The answer
is found in a property of the global time cutoff measure which is
outside our experience, but which is undeniably a feature of
these measures: the youngness bias.  There will be many copies of
me scattered around the multiverse, all trying the same sleeping
experiment that I chose to do.  If $Z$ is large, there will be
many more in the first hour of the experiment than in any later
hour, so the first hour is strongly weighted in the multiversal
averages.  We can check this quantitatively by writing down the
probability distribution for copies of me in the sample spacetime
region, carrying out this same experiment, as a function of the
time $t$ since the start of the experiment (when the Higgs mass
is measured).  The probability falls with $t$ as $e^{- \lambda
t}$, but there are also corrections due to sleep and death. 
Thus, if a random copy of me during this experiment were found in
the multiverse, the probability distribution for $t$ would be
\beq
   p(t) = e^{-\lambda t} \times \lcurl \begin{array}{cl}
            \half A & \hbox{if\ } \Delta t_{\rm short} < t
                             < \Delta t_{\rm long}\\
              A  & \hbox{if\ } \Delta t_{\rm long} < t
                   < t_{\rm max} + \Delta t_{\rm short}\\
            \half A & \hbox{if\ } t_{\rm max} + \Delta t_{\rm short}
                   < t < t_{\rm max} + \Delta t_{\rm long}\ ,\\
          \end{array}\right.
\label{eq:p(t)}
\eeq
where the normalization constant $A$ is determined by $\int p(t)
{\rm d} t = 1$, which gives
\beq
   A = {2 \lambda \over \left( e^{- \lambda \Delta t_{\rm short}}
     + e^{- \lambda \Delta t_{\rm long}} \right) \left( 1 - e^{-
     \lambda t_{\rm max}} \right)} \ .
\label{eq:A}
\eeq
By combining Eqs.~(\ref{eq:equalt}), (\ref{eq:p(t)}), and
(\ref{eq:A}), one finds that the weighted average over time of
$P(1;t,t)$ is given by
\beq
   \expect{P(1;t,t)} \equiv \int P(1;t,t) \, p(t) \, {\rm d} t = {Z
     \over Z+1} \ ,
\label{eq:weighted}
\eeq
recovering the result from Eqs.~(\ref{eq:wakeup}) and
(\ref{eq:wakeup+t}). Thus, the surprising effect of the measure
is not that it can affect the Higgs mass, but rather that if $Z
\gg 1$, the strong youngness bias can make the first hour more
important to the weighted average probability than the entire
rest of my life. Since the relative weighting of time given by
$p(t)$ was crucial in turning the unsurprising expression for
$P(1;t,t)$ of Eq.~(\ref{eq:equalt}) into the surprising $Z:1$
odds described by Eq.~(\ref{eq:weighted}), we should ask to what
extent $p(t)$ is meaningful to individuals living in a multiverse
described by a global time cutoff measure with a large
exponential expansion rate.  If I live in such a world, do I have
to believe that earlier times in my life are much more important
than later times?  Our view is that the $e^{- \lambda t}$ factor
in $p(t)$ reflects the fact that there are more young copies of
me in the multiverse than there are older copies of me.  But the
bias toward younger times affects me directly only if I somehow
become uncertain about my age, in which case I need to use the
probability distribution to determine what my age is likely to
be.  The G-V thought experiment is an example of this, where the
sleeper does not know when he wakes up whether one minute or one
hour has elapsed.  Another example is the cosmological youngness
bias, as discussed in Sec.~\ref{sec:youngness}, which rules out
proper time measure because of the prediction it makes, for
example, for the cosmic microwave background temperature.  In
this case the youngness bias combines with our uncertainty about
how much time is needed for a civilization to evolve from the big
bang, producing a striking prediction.  But in the mathematical
limit interpretation that we advocate, the youngness bias does
not mean that time will end, and it also does not mean that I
should expect to die young.  The probability that I will be alive
or dead at age 90 is uninfluenced by the youngness bias. Thus, I
am free to make my own choices about the relative importance of
the different stages in my life.  Issues of deferred
gratification are not controlled by the measure, but will remain
a subject to be studied by psychologists and sociologists.  While
it is true that the late stages of my life will contribute little
to multiversal averages, I can still value them as much as I
choose.

\section{Conclusion}

BFLR concluded their paper by stating that the deduction that
time can end can be avoided only by rejecting at least one of
three propositions:

\begin{itemize}

\item[(1)] Probabilities in a finite universe are given by
relative frequencies of events or histories.

\item[(2)] Probabilities in an infinite universe are defined by a
geometric cutoff.

\item[(3)] The Universe is eternally inflating.

\end{itemize}

We believe that there is a fourth possibility, which these
authors have passed over: there is a mathematically well-defined
way of defining probabilities without imposing an end of time,
which can be adopted without rejecting any of the three
statements above.  In this paper we have tried to describe how
this works.  In summary, the procedure begins by constructing a
classical (stochastic) mathematical model of the multiverse as an
infinite system, defined on a fine-grained lattice that grows
exponentially like the multiverse itself.  The model is defined
by choosing a probability distribution for the field values at
the initial time step, and then giving an update rule that
determines the probabilities for the $n^{\rm th}$ time step in
terms of the previous ones.  We believe that such a model can be
as well-defined as the properties of the integers.  Since the
model is infinite from the start, the danger that time might end
does not seem to be present.  Following the standard procedure
for a global time cutoff, one then defines a sample spacetime
region in the model, which grows without limit as some final
cutoff time $\tau_c$ approaches infinity.  The relative
probability between any two types of stories is then defined as
the ratio of the number of occurrences in the sample spacetime
region, in the limit as $\tau_c \to \infty$. As far as we can
tell, this system is rigorously mathematically consistent. 
However, we do have to admit that it has counter-intuitive
properties.  In particular, this system predicts that if the
outcome of some experiment is reported with a time delay, where
the length of the delay depends on the result, then the
observation of the reports will be biased.  The probability of
observing the report with the shortest time delay will be higher
than the probability that this result occurred.  For measures
such as scale-factor cutoff measure, this effect would be far too
small to observe, but it will be present in principle.  While it
is easy to see how the time-delay bias arises in the global time
cutoff calculation, some physicists might still find it
offensively counter-intuitive.  Since we are not arguing that
this measure is correct --- only that it is consistent --- it is
certainly appropriate for anyone who finds this approach
counter-intuitive to look for other solutions to the measure
problem.  Counter-intuitiveness is a subjective judgment.  We,
however, feel that the transition from conventional probability
to probabilities in the multiverse is a sufficiently large step
that we should not expect all of our conventional intuition to
carry over.  Finally, we should clarify that the probability
measure proposed here is only a prescription.  That is, we are
describing a way to define probabilities, but we do not know of
any physical mechanism that might cause it to be the correct
definition to use for making predictions.  On this issue, the
end-of-time approach has a possible advantage.  If time really
does end, then the spacetime becomes finite, and probabilities
can be unambiguously defined by counting.  Thus, if one wants to
not merely have a prescription for defining probabilities, but to
also understand what makes it the right prescription, then the
end-of-time hypothesis is one way to achieve this goal.

\begin{acknowledgments}
Without suggesting that any of these authors agree with our
conclusions, the authors would like to thank the following people
for very helpful discussions: Tom Banks, Raphael Bousso, Adam
Brown, Ben Freivogel, Jenny Guth, Larry Guth, Sam Gutmann, Daniel
Harlow, Matthew Kleban, Stefan Leichenauer, Andrei Linde,
Mahdiyar Noorbala, Ken Olum, Don Page, Vladimir Rosenhaus,
Michael Salem, Delia Schwartz-Perlov, Stephen Shenker, Ben
Shlaer, David Spiegel, Douglas Stanford, Lenny Susskind, and
Alexander Vilenkin.  The authors also wish to acknowledge the
hospitality of the Foundational Questions Institute (FQXi), which
sponsored the conference at which this work began, and the
Perimeter Institute, where much of the work in completing the
paper was carried out.  AHG is supported in part by the DOE under
contract No.\ DE-FG02-05ER41360.  VV is supported in part by NSF
Grant No.\ 0756174.
\end{acknowledgments}

\appendix

\section{Cloning and Time-Dependent Probabilities}

In Sec.~\ref{sec:definition} we defined the probability for the
outcome of any experiment in terms of the counting of stories in
the sample spacetime region, and in Sec.~\ref{sec:stepbystep} we
showed how this definition gives rise to a time-delay bias: if
the outcome of some experiment is reported with a time delay,
where the length of the delay depends on the result, then the
observation of the reports will be biased.  The probability of
observing the report with the shortest time delay will be higher
than the probability that this result actually occurred.
In the examples discussed in this paper we assumed that the local
time variable $t$ is related to the global cutoff time variable
$\tau$ in such a way that ${\rm d} t / {\rm d} \tau \approx
\hbox{const}$ during any experiment, but this need not be the
case.  In particular, if we are using scale-factor cutoff measure
to predict the cosmological constant $\Lambda$, then ${\rm d} t /
{\rm d} \tau$ would depend on $\Lambda$ and could not be treated
as a constant.  In that case one would find that the probability
of different outcomes could change with the passage of time, over
cosmological time scales, even if there is no reporting delay.
The discussion in the paper about how time-dependent
probabilities arise is self-contained and complete, but since the
notion that probabilities can change with time is so contrary to
our experience, we will use this appendix to show in simple
examples how this can happen.  The key element is cloning, where
by cloning we mean the creation of indistinguishable copies of an
object.  We use the same word whether the creation of the copy is
intentional, as happens sometimes in thought experiments, or
accidental, as we expect in the multiverse when stories are
repeated over and over again in the eternal growth of the system.
One simple example of how cloning can lead to time-dependent
probabilities is a variant of the often-discussed ``Sleeping
Beauty'' problem \cite{Zuboff:1990, Elga:2000, Bostrom:2007}. 
Suppose that a fair coin is flipped, and is placed under a hat on
a table in front of Sleeping Beauty.  The experimenter looks at
the coin, but does not show it to Sleeping Beauty.  There is also
a clock on the table.  Sleeping Beauty is left alone in the room,
but is told that when the clock reads 12:00, if the coin is a
head, an identical copy would be created of her and the entire
room.  If the coin is a tail, nothing is done.  Before the clock
reads 12:00, the probability that the coin is a head is clearly
50\%, since it is a fair coin.  After 12:00, what is it?  The
literature on the sleeping beauty problem includes papers
supporting a wide variety of approaches, but here we follow a
counting procedure that we find straightforward, and analogous to
the counting procedure we are proposing for the multiverse.  If
the experiment were repeated $N$ times, the expectation value for
the number of heads and tails are each $N/2$.  If $N$ is chosen
to be a large number, the standard deviation would be relatively
small, so we can expect $N/2$ to be a good approximation to the
actual number of heads and tails.  Since Sleeping Beauty is
cloned in the case of heads, there would be $N$ Sleeping Beauties
who would find that the coin is a head, and $N/2$ who would find
a tail.  Thus, the probability is 2/3 that the coin is a head. 
Even though the coin has already been flipped, Sleeping Beauty
would conclude as the clock reaches 12:00 that the probability of
the coin being a head suddenly changes from 1/2 to 2/3.  This may
not look much like the multiverse, but it illustrates the
principle that probabilities can change with time. 

While the cloning of a person is technologically out of sight,
the cloning of a computer is easy.  Starting with two computers
with matching hardware, one can arrange for the memory and hard
drives of the two computers to match byte by byte. Computers can
also be switched on and off, with no ambiguity about whether they
can detect the passage of time when they are off. So, to
continue, we will imagine that our subjects are computers.
\begin{figure}[htbp]
   \begin{center}
   \includegraphics{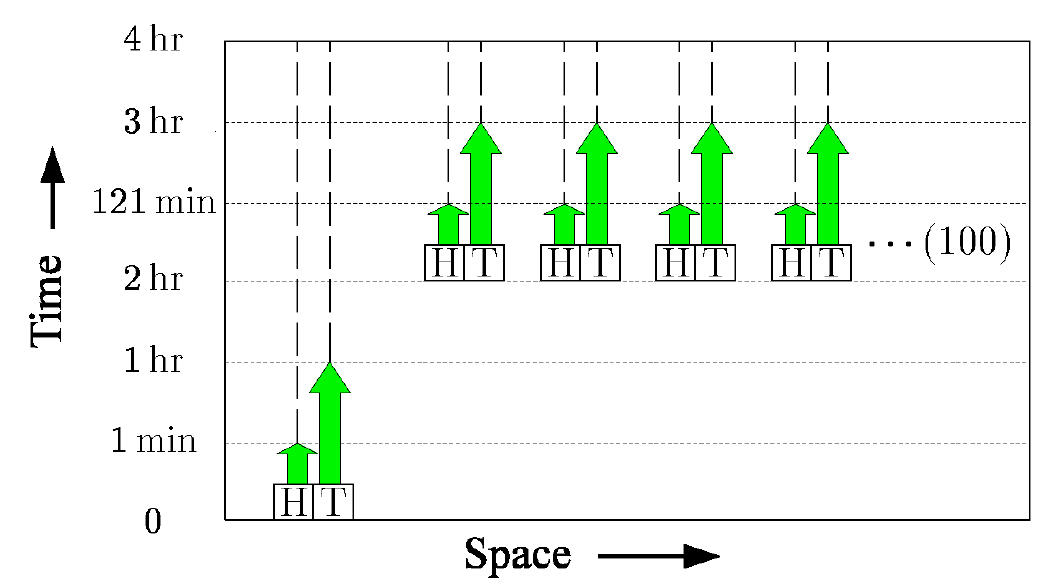}
   \caption{The cloning of a G--V thought experiment.}
   \label{fig:cloning1}
   \end{center}
   \end{figure}
To start,
consider the sequence of events described in Fig.\
\ref{fig:cloning1}.  This is still not the multiverse model, but 
it is a useful steppingstone.  At $t=0$, two computers are
deployed.  The computers are identical, except that the first is
a Head Computer, with an H engraved on its case, and the second
is a Tail Computer, with a T engraved on its case.  Neither
computer can read its engraving, but they are programmed to
compute the probability that it is an H\relax.  The Head Computer
is turned on after one minute, but the Tail Computer is turned on
after 1 hour.  The computers have internal clocks; when each
computer is turned on it is set to the current time, so once the
computers are both running they are perfect clones of each other,
matching at every instant of time.  The computer programs are
written with knowledge of the procedures for the experiment, but
the same program runs on both computers, so it cannot know if it
is running on an $H$ computer or a $T$ Computer.  Between 1 min
and 1 hr, therefore, the program concludes that it is 100\%
likely to be the $H$ Computer, since the $T$ Computer is
scheduled to be off.  At $t=1$ hr, the program knows that the $T$
computer is switched on, and also that it has no way to determine
if it is running on the $H$ computer or the $T$ computer.  It
therefore announces that the probability is now only 50\% that it
is the $H$ computer.  At $t=2$ hr, 100 more pairs of $H$ \& $T$
computers are deployed, all identical to the originals. The $H$
computers are switched on at $t=121$ min, while the $T$ computers
are not switched on until $t=3$ hr.  The internal clocks are
always set to the current time, so all running computers are
perfect clones of each other.  Now the computers announce at
$t=121$ min that the probability of being an $H$ computer is
101/102.  At $t=3$ hr the probability of $H$ returns to 50\%. 
The probability is simply determined by the population of
computers that are currently running, which can shift in
arbitrary ways.  In our example the computers are always deployed
in $H$--$T$ pairs, but they are turned on at different times,
resulting in nonuniform probabilities.  Our picture of the
multiverse differs in an important way from the above example, in
that time is treated differently. In the multiverse we imagine
that a global time coordinate can be defined, as we did in our
description of a lattice simulation, but observers have no way of
accessing this coordinate.  The statistics of the lattice
simulation are regularized by selecting a sample spacetime
region, which can become larger and larger as the final cutoff
time $\tau=\tau_c$ is taken to infinity.  For any value of
$\tau_c$, as the limit is taken, we can determine the predictions
for any experiment by imagining that the experiment we are
carrying out is equally likely to be any copy of the experiment
that we can find in the sample spacetime region.  There is no
notion of the current time in the sample spacetime region, but
there is a cutoff time $\tau=\tau_c$, which plays a very similar
role.
\begin{figure}[htbp]
   \begin{center}
   \includegraphics{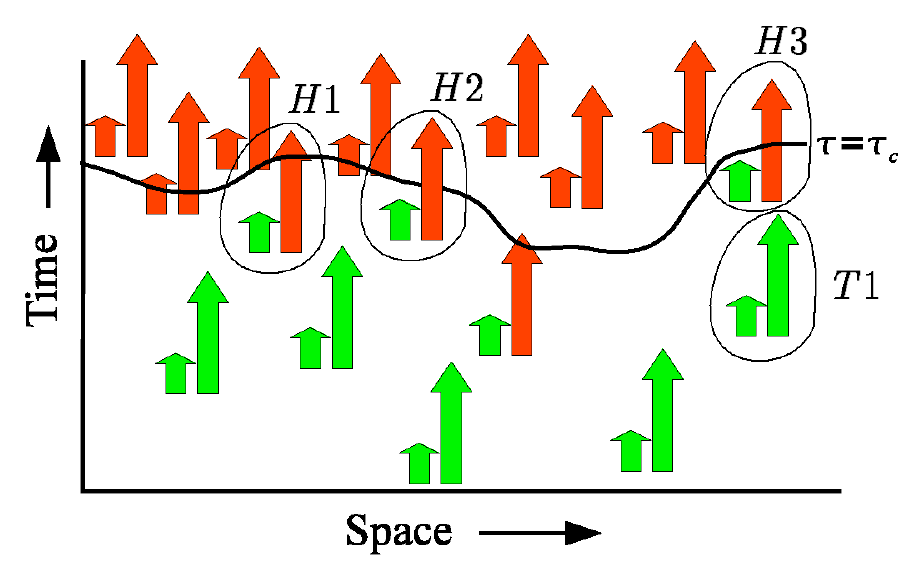}
   \caption{The cloning of the G--V thought experiment in the multiverse.}
   \label{fig:cloning2}
   \end{center}
   \end{figure}
Suppose we consider the G--V thought experiment, which was
diagrammed in the multiverse as Fig.~\ref{fig:paradox}.  We
have redrawn the diagram with some extra annotation as
Fig.~\ref{fig:cloning2}.  If we wish to consider the possibility
that I might be the Tail Subject, then one of the possible
instances would be the one marked $T1$.  But for any instance,
such as $T1$, for which the Tail Subject wake-up fits under the
cutoff, there will be many copies of the experiment that begin
later, such as $H1$, $H2$, and $H3$.  Each copy involves both a
Head Subject and a Tail Subject, but in most of these copies the
awakening of the Tail Subject will not occur under the cutoff. 
Thus, when we take into account both the cloning (i.e., the
appearance of new copies) and the cutoff, we can see how the
global time cutoff measure predicts more Head Subject wake-ups
than Tail Subject wake-ups.  (Note that the use of the cutoff is
not optional: it is the defining property of the global time
cutoff measure.)

\end{document}